\documentclass[a4paper,fleqn,usenatbib]{mnras}

\usepackage{newtxtext,newtxmath}

\usepackage[T1]{fontenc}
\usepackage{ae,aecompl}

\usepackage{graphicx}	
\usepackage{amsmath}	
\usepackage{amssymb}	

\newcommand{\kms}{\mbox{km\,s$^{-1}$}}

\newcommand{\etal}{\hbox{et al.}} 
\newcommand{\vlsr}{\hbox{v$_{\tiny{\textrm{LSR}}}$}}

\newcommand{\Tpeak}{\hbox{T$_{\tiny{\textrm{Peak}}}$}}

\newcommand{\RXJ}{\hbox{RX\,J1713.7$-$3946}}

\newcommand{\Jsto}{\hbox{HESS\,J1731$-$347}}
\newcommand{\Jstn}{\hbox{HESS\,J1729$-$345}}

\newcommand{\cmsqr}{\hbox{cm$^{-2}$}}

\newcommand{\Msun}{\hbox{M$_{\odot}$}}
\newcommand{\miriad}{\hbox{\sc Miriad}}

\newcommand{\livedata}{\hbox{\sc Livedata}}
\newcommand{\gridzilla}{\hbox{\sc Gridzilla}}

\title[Probing $\Jsto$ with CO and CS]{Probing The Local Environment of the Supernova Remnant {\Jsto} with CO and CS Observations}

\author[Maxted \etal]{N. Maxted,$^1$\thanks{E-mail: n.maxted@unsw.edu.au}
M. Burton,$^{1,2}$
C. Braiding,$^{1}$
G. Rowell,$^{3}$
H. Sano,$^{4}$
F. Voisin,$^{3}$
\newauthor M. Capasso,$^{5}$
G. P\"uhlhofer,$^{5}$
and  Y. Fukui$^{4}$ 
\\
$^1$School of Physics, The University of New South Wales, Sydney, 2052, Australia\\
$^2$Armagh Observatory and Planetarium, College Hill, Armagh, BT61 9DG, Northern Ireland, United Kingdom\\ 
$^3$School of Physical Sciences, The University of Adelaide, Adelaide, 5005,  Australia\\ 
$^4$Department of Astrophysics, Nagoya University, Furocho, Chikusa-ku, Nagoya, Aichi, 464-8602, Japan\\ 
$^5$Institut f\"ur Astronomie und Astrophysik, Universit\"at T\"ubingen, T\"ubingen, 72076, Germany 
}

\date{Accepted XXX. Received YYY; in original form ZZZ}             

\pubyear{2017}

\begin{document}
\label{firstpage}
\pagerange{\pageref{firstpage}--\pageref{lastpage}}
\maketitle

\begin{abstract}
The shell-type supernova remnant $\Jsto$ emits TeV gamma-rays, and is a key object for the study of the cosmic ray acceleration potential of supernova remnants. 
We use 0.5-1\,arcminute Mopra CO/CS(1-0) data in conjunction with HI data to calculate column densities towards the $\Jsto$ region. 
We trace gas within at least four Galactic arms, typically tracing total (atomic+molecular) line-of-sight H column densities of 2-3$\times$10$^{22}$\,cm$^{-2}$. Assuming standard X-factor values and that most of the HI/CO emission seen towards $\Jsto$ is on the near-side of the Galaxy, X-ray absorption column densities are consistent with HI+CO-derived column densities foreground to, but not beyond, the Scutum-Crux Galactic arm, suggesting a kinematic distance of $\sim$3.2\,kpc for $\Jsto$. 
At this kinematic distance, we also find dense, infrared-dark gas traced by CS(1-0) emission coincident with the north of $\Jsto$, the nearby HII region G353.43$-$0.37 and the nearby unidentified gamma-ray source $\Jstn$. This dense gas lends weight to the idea that $\Jstn$ and $\Jsto$ are connected, perhaps via escaping cosmic-rays.
\end{abstract}





\section{Introduction}\label{sec:intro}
\begin{figure}
\includegraphics[width=0.48\textwidth, angle=0]{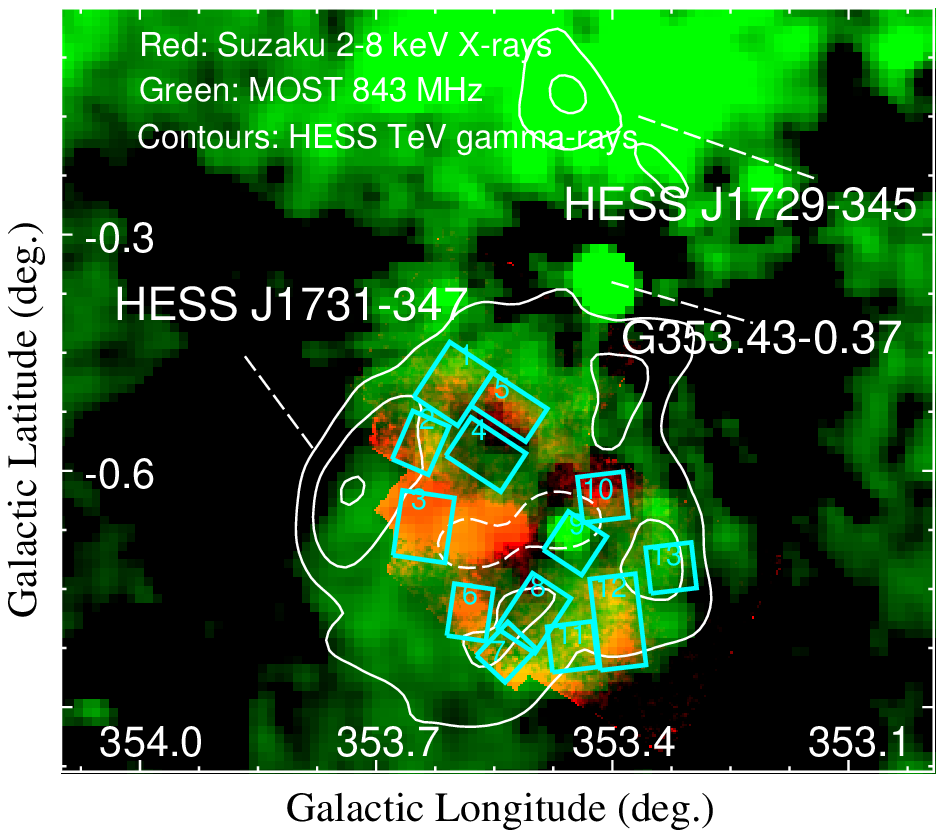}
\caption{Suzaku 2-8\,keV X-ray emission over a limited field towards $\Jsto$ (red) with boxes indicating regions used for X-ray absorption analyses \citep{Bamba:2012}. The image also features 843\,MHz radio continuum \citep{Green:1999} emission (green) and HESS $>$1\,TeV 4, 6 and 8$\sigma$ \citep{Abramowski:2011} contours (where the thin-dashed contours indicate a central 4$\sigma$ void). The position of SNR $\Jsto$, unidentified gamma-ray source $\Jstn$ and the HII region G353.43$-$0.37 are indicated. \label{fig:1}}
\end{figure}

$\Jsto$ is a TeV gamma-ray source \citep{Aharonian:2008,Abramowski:2011} believed to be associated with the shell-type supernova remnant (SNR) G353.6$-$0.7 \citep{Tian:2008}, while the nearby gamma-ray source $\Jstn$ was more recently detected in gamma-rays and is yet to be identified \citep{Abramowski:2011}. These objects, which are displayed in Figure\,\ref{fig:1}, are both candidates in the search for cosmic-ray hadron (hereafter `CR' refers to cosmic ray hadrons unless otherwise stated) sources within the Galaxy. 

Much progress has been made in identifying individual Galactic CR sources. To explain the Galactic Centre TeV gamma-ray distribution, it has been proposed that Sagittarius A* is a past accelerator of CRs up to PeV energies \citep{Abramowski:2016nature}, while GeV gamma-ray spectral studies have revealed `smoking-gun' CR sources SNRs W44 and IC\,443 \citep{Ackermann:2013}. Multi-wavelength studies expose other supernova remnants as probable CR sources \citep[e.g. W28, W49B and HESS\,J1640$-$465][]{Aharonian:w28,Nicholas:2011,Nicholas:2012,Maxted:2016a,Abdalla:2016w49,Lau:2017}. Even when accounting for these objects, the origin of the total flux of observed Galactic CRs still remains an open question. In the search for current $>100$\,TeV CR accelerators, young SNRs (e.g. SNR\,$\RXJ$, \citealt{Fukui:2012,Gabici:2014}) are a key target. 

With a gamma-ray flux of $\sim$10\% the Crab flux above 1\,TeV, $\Jsto$ was one of the brightest TeV objects without an identified counterpart until \citet{Tian:2008} discovered corresponding radio continuum and X-ray emission counterparts, both displayed in Figure\,\ref{fig:1}. The radio emission is non-thermal and shell-like in structure, and soft X-ray emission measured with ROSAT is only seen from the hemisphere of the TeV-SNR farthest from the Galactic Plane, suggesting absorption by a foreground molecular cloud in the hemisphere nearest to the plane (see Section\,\ref{sec:Distance}). The authors proposed a potential association of the SNR with molecular gas connected with the HII region G353.43$-$0.37, then calculated a radiative SNR age of $\sim$2.7$\times$10$^4$\,yr. This was noted to be consistent with the non-thermal X-ray flux relative to the TeV gamma-ray flux according to \citet{Yamazaki:2006}, but later studies suggest a younger age \citep[e.g. $\sim$10$^3$\,yr,][]{Acero:2015,Cui:2016}. Follow-up studies incorporated XMM-Newton X-ray observations to study the compact central object (CCO), XMMS\,J173203$-$344518 \citep{Tian:2010}, which had a detected pulsation that was later shown to be tenuous \citep{Halpern:2010a}, then absent \citep{Halpern:2010b}. A Suzaku study found hard and featureless non-thermal X-ray emission and confirmed that the foreground X-ray absorption column density varies across the SNR shell \citep{Bamba:2012}, suggesting that the SNR is behind molecular gas of the Scutum-Crux arm.

The CCO has been the laboratory for a number of neutron star studies \cite[e.g.][]{Klochkov:2013,Ofengeim:2015,Klochkov:2015}, and recent work by \citet{Doroshenko:2016} suggests that the $\Jsto$  candidate progenitor may be part of a binary system.

$\Jsto$ shares many qualities in common with young ($\sim$10$^3$\,yr) shell-type SNRs - $\RXJ$, RX\,J0852.0$-$4622, RCW\,86 and SN\,1006, suggesting a common TeV gamma-ray emission mechanism \citep{Acero:2015}. Like these remnants, observations of the $\Jsto$ non-thermal X-ray emission has shown the remnant to be accelerating electrons (leptons) \citep{Tian:2008,Tian:2010,Bamba:2012}, possibly via a diffusive-shock mechanism \citep[see e.g.][]{Bell:1978}. It logically follows that $\Jsto$ may also accelerate hadrons beyond TeV energies, such that part of the observed gamma-ray shell or nearby gamma-ray sources may be a by-product of CRs interacting with matter, producing neutral pions which decay into gamma-ray photons (hadronic scenario). If such a hadronic mechanism was demonstrated to be occurring towards HESS\,J1731$-$347, the SNR would be shown to be a CR accelerator. This scenario is confused by presence of a competing mechanism for gamma-ray production, inverse Compton scattering, whereby the same population of X-ray-emitting electrons up-scatters stellar/CMB background photons to create the observed gamma-ray emission (leptonic production). 

Upper limits placed on GeV emission \citep{Yang:2014}, at first glance, do not seem to support a hadronic scenario due to the non-detection of a so-called `pion-bump' in the GeV-part of high-energy spectrum, but this does not rule out CR acceleration within $\Jsto$. CRs may contribute to a component of the gamma-ray spectrum, a scenario that becomes more likely if localised regions containing large magnetic fields and dense clumps of gas are present. In such cases, higher energy CRs more successfully access CR target material than lower energies, resulting in a hadronic gamma-ray spectrum that lacks a characteristic pion-bump \citep{Fukuda:2014}, as has been suggested for SNR\,$\RXJ$ \citep{Gabici:2009,Zirakashvili:2010,Inoue:2012,Fukui:2012,Maxted:2012,Gabici:2014}. \citet{Fukuda:2014} find a correlation between proton density (at a kinematic distance of 5.2-6\,kpc) and TeV gamma-ray emission across $\Jsto$ leading them to advocate for a $\Jsto$ hadron-dominated ($\sim$80\%) gamma-ray emission scenario. No thermal X-ray emission from the deposition of the $\Jsto$ shock energy into diffuse gas is observed, but this is consistent with the scenario of a SNR evolving into an inhomogeneous clumpy medium, like that of a wind-blown bubble \citep{Inoue:2012}. 

Modelling by \citet{Cui:2016} does not assume the existence of the hadronic component suggested by the TeV-gas correlation found by \citet{Fukuda:2014}, and uses the measured gamma-ray spectrum to favour a lepton-dominated scenario arising from a 20-25\,{\Msun} progenitor for $\Jsto$. Cloud geometry was then invoked as a possible explanation for a hadronic origin for $\Jstn$, consistent with CO emission from gas at 3.2\,kpc. A key aspect that allows such a hadronic feature from `runaway' CRs for $\Jsto$ is the larger proton interaction mean-free path as compared to high energy electrons which quickly lose energy via the synchrotron process, meaning CRs are able to produce gamma-rays via pion production/decay farther away from the SNR shell than electrons are able to produce gamma-rays via bremsstrahlung or inverse-Compton. A preliminary detection of a TeV gamma-ray extension between $\Jsto$ and $\Jstn$ is suggestive of runaway CRs from $\Jsto$ \citep{Capasso:2016}. Either a clear case of runaway CRs outside the boundary of $\Jsto$ or the detection of hadronic gamma-ray components within a seemingly lepton-dominated $\Jsto$ gamma-ray spectrum would expose the remnant as a CR accelerator.

\subsection{Distance}\label{sec:Distance}
\citet{Fukuda:2014} argued that the $\Jsto$ SNR is associated with a void in HI-traced atomic gas on the near side of the 3\,kpc-expanding arm at a kinematic distance of 5.2-6\,kpc (line of sight velocity $\sim - 85$\,km\,s$^{-1}$). This scenario is consistent with previous work comparing the X-ray absorption column density distribution towards $\Jsto$ with CO and HI-derived column densities to conclude that the SNR lies behind the Scutum-Crux arm gas, which is at a Galactic kinematic distance of $\sim$3.2$\pm$0.8\,kpc ($\vlsr \sim -$18\,$\kms$) \citep{Tian:2010}, and is at a similar kinematic distance to the HII region G353.43$-$0.37 \citep{Bronfman:1996,Quireza:2006}. 
This distance is consistent with new work by \citet{Doroshenko:2017}, who find a spatial pattern of X-ray absorption in XMM-Newton data that matches the 3.2\,kpc kinematic distance gas, reinforcing previous lower limits.


The gas analysis in this paper offers higher spectral and spatial resolution CO(1-0) data than previous molecular surveys towards $\Jsto$ and $\Jstn$. A key component of this investigation is to utilise Mopra CO data in conjunction with SGPS HI data \citep{McClure:2005} and X-ray absorption column densities \citep{Bamba:2012} to derive a $\Jsto$ kinematic distance (see Section\,\ref{sec:Xray}). Additionally, a new multi-wavelength morphological study of the coincident CO-traced molecular gas is undertaken. $^{13}$CO, CS and C$^{34}$S complement the data-set by probing dense gas that may contribute to a hadronic gamma-ray emission in the region.

\section{Observations}
\subsection{The Mopra CO Galactic Plane Survey at 3\,mm}
Spectral data of the J$=$1-0 transition of CO isotopologues was taken as part of the Mopra Galactic Plane CO Survey\footnote{http://phys.unsw.edu.au/mopraco/} \citep[see][ for the full data-reduction description]{Burton:2013}, which will be made public in stages \citep[see][ for the first data release]{Braiding:2015}\footnote{see www.mopra.org/data/}. 
Four isotopologues are targeted in this survey: CO, $^{13}$CO, C$^{18}$O and C$^{17}$O. 
The former two are the most abundant, so are exploited in this paper. 

The Mopra Galactic Plane CO Survey data-set has an angular resolution and a velocity resolution of 35$^{\prime\prime}$ and 0.1\,kms$^{-1}$, respectively, across 8 4096-channel dual-polarisation bands. After data reduction, the longitude range presented in this paper has a full velocity coverage of $-$185\,kms$^{-1}$ $<$v$_{LSR}<+$115\,kms$^{-1}$, with negative and positive velocity extensions of  $\sim$60\,kms$^{-1}$ in the north and south of the field, respectively. The extended beam efficiency used to scale-up the antenna temperature of Mopra at 115\,GHz is 0.55 \citep{Ladd:2005}. 
Typical $\sim$1$\sigma$ noise values are 0.8\,K\,channel$^{-1}$. 
Full details of the CO data reduction are presented in \citet{Braiding:2015}.

\begin{figure*}
\includegraphics[width=1.02\textwidth, angle=0]{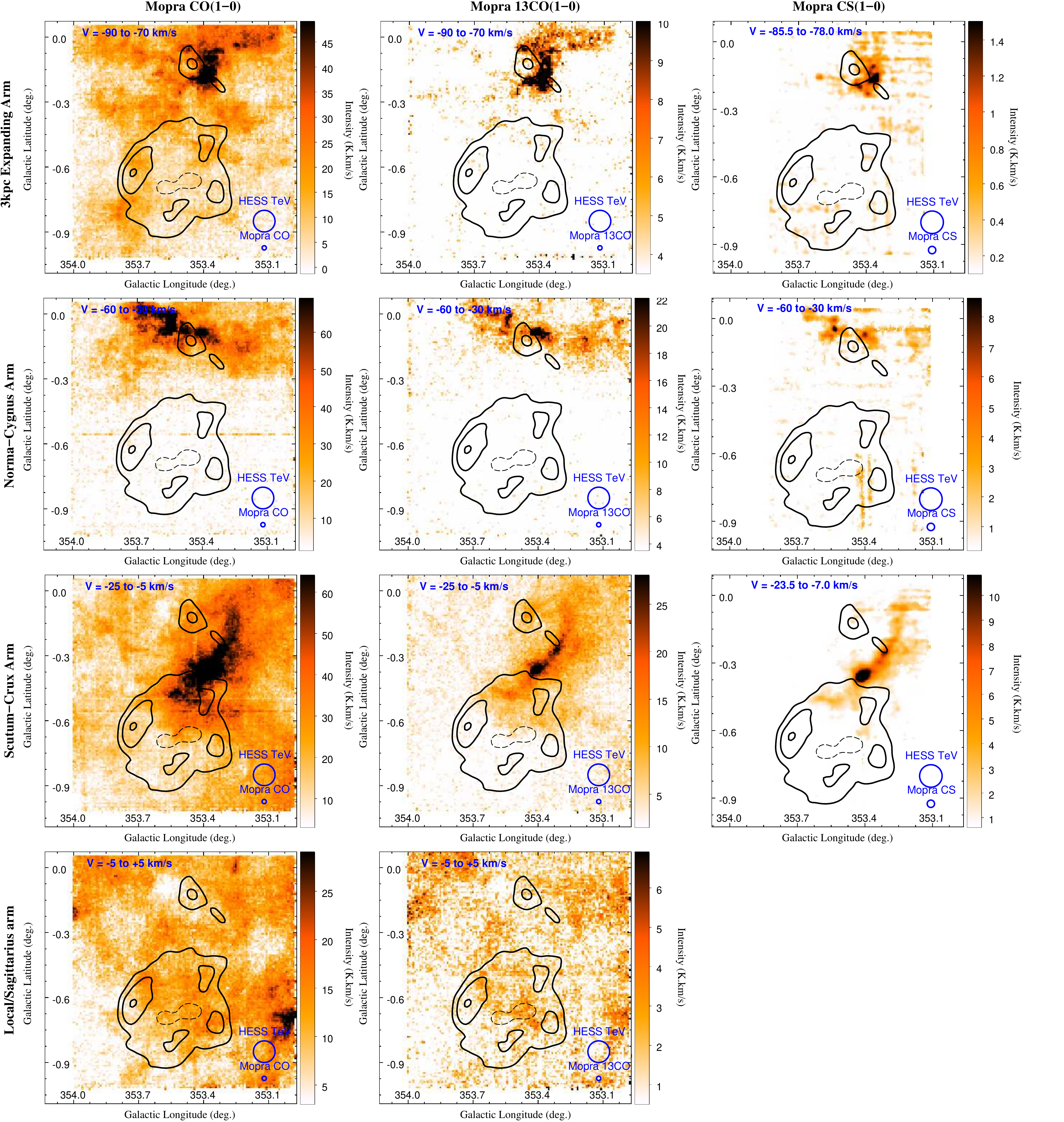}
\caption{Velocity integrated CO(1-0) (left), $^{13}$CO(1-0) (middle) and CS(1-0) (right) emission towards $\Jsto$ and $\Jstn$ for velocity ranges corresponding to four Galactic arms, the 3kpc-Expanding ($-$90 to $-$70\,$\kms$, top), Norma-Cygnus ($-$60 to $-$30\,$\kms$), Scutum-Crux ($-$25 to $-$5\,$\kms$) and Sagittarius ($-$5 to $+$5\,$\kms$, bottom) arms. HESS 4, 6 and 8$\sigma$ $>$1\,TeV gamma-ray significance contours (thick black) are overlaid \citep{Abramowski:2011}. Thin-dashed contours indicate a central 4$\sigma$ void. Precise velocity range and instrument beam FWHM of each image are indicated. 
\label{fig:mom0s_1}}
\end{figure*}

\begin{figure}
\includegraphics[width=0.49\textwidth, angle=0]{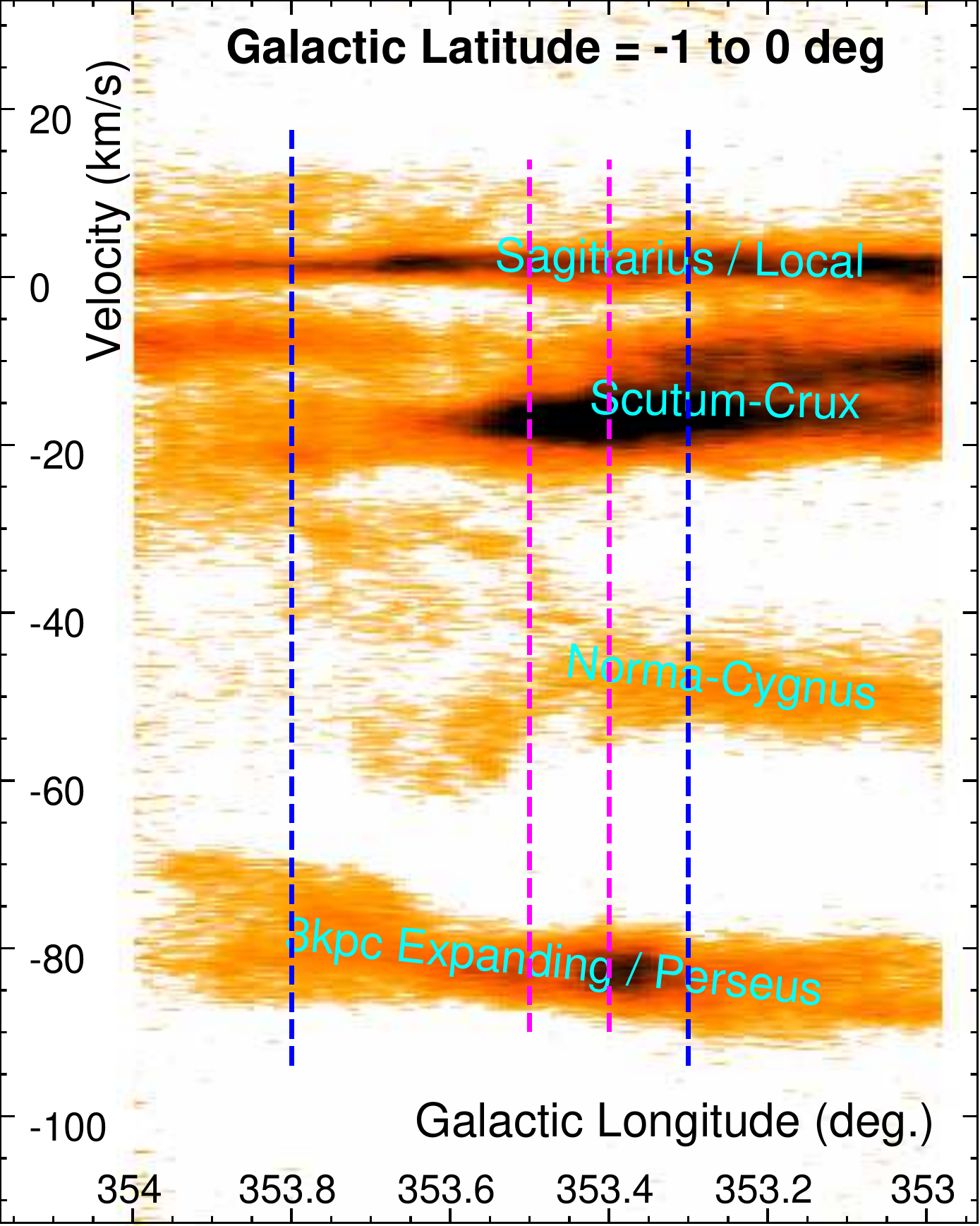}
\caption{A plot of Mopra CO(1-0) emission as a function of line of sight velocity and Galactic Longitude. CO emission has been spatially integrated between $-$1 and 0$^{\circ}$. Blue and pink dotted lines indicate the longitudinal extent of $\Jsto$ and $\Jstn$, respectively. \label{fig:PVplot}}
\end{figure}

CO(1-0) and $^{13}$CO(1-0) integrated intensity images are shown alongside CS(1-0) images in Figure\,\ref{fig:mom0s_1}. Velocity-integration ranges were chosen to highlight four Galactic arm structures illustrated in Figure\,\ref{fig:PVplot}.

\subsection{The Mopra 7\,mm Survey Targeting CS}
The 7\,mm Mopra survey of $\Jsto$ and $\Jstn$ is comprised of a mixture of 26 partial and full OTF (on the fly) maps towards 7 different 21$^{\prime}\times$21$^{\prime}$ fields, which were taken between March 2011 and May 2014. 

The Mopra spectrometer, MOPS, is capable of recording sixteen tunable, 4096-channel (137.5\,MHz) bands simultaneously when in `zoom' mode, as used here. The data were co-added to produce 16 data-cubes with 2 spatial (long/lat) and 1 spectral (velocity) dimension at 7\,mm wavelengths. 
Typical $1\sigma$ antenna noise values are $\sim$0.1\,K for the CS(1-0) transition at 48.990957\,GHz, but the exposure varies over the mapped field leading to some variation of this value on a scale of $\sim$50\%.

Mopra mapping data have a cycle time of 2.0\,s and the spacing between scan rows is 26$^{\prime\prime}$. The velocity resolution of the 7\,mm zoom-mode data is $\sim$0.2\,kms$^{-1}$. The beam FWHM and the pointing accuracy of Mopra at 7\,mm are 59$\pm$2$^{\prime\prime}$ and $\sim$6$^{\prime\prime}$, respectively. Beam efficiencies presented in \citet{Urquhart:2010} were applied to convert antenna intensity into main-beam intensity (0.56 for extended CS(1-0) emission at 49\,GHz).

Data were reduced using the software packages $\livedata$, $\gridzilla$ \citep{Gooch:1996} and $\miriad$ \citep{Sault:1995}. $\livedata$ software performed a bandpass calibration on each row, using the preceding off-scan as a reference, then applied a linear fit to the baseline. $\gridzilla$ software weighted the data by the system temperature, smoothed the data using a gaussian of FWHM equal to the Mopra beam ($\sim$1$^\prime$, extending outwards 3$^\prime$) and averaged the data from individual scans into data cubes. The resultant cubes are in fits-file format and have pixels of size ($\Delta\,x,\Delta\,y,\Delta\,z$)=(15\arcsec, 15\arcsec, 0.2\,\kms). Mopra data fits files, and ds9 region/contour files presented in this paper will be made publicly available\footnote{www.physics.adelaide.edu.au/astrophysics/MopraGam/}.

\subsection{21\,cm SGPS HI data}
Publicly-available 21\,cm HI data from the Southern Galactic Plane Survey \citep[see][ for details]{McClure:2005} was exploited in this analysis. The HI data were taken with the Parkes single-dish telescope and the ATCA interferometer, and the resultant fits-file probed atomic gas at an angular resolution and a velocity resolution of $\sim$2$^{\prime}$ and $\sim$0.8\,$\kms$, respectively. The HI data used in this analysis have had radio-continuum sources removed.

\section{Spectral Line Analyses}\label{sec:analysis}
\begin{figure*}
\includegraphics[width=1.02\textwidth, angle=0]{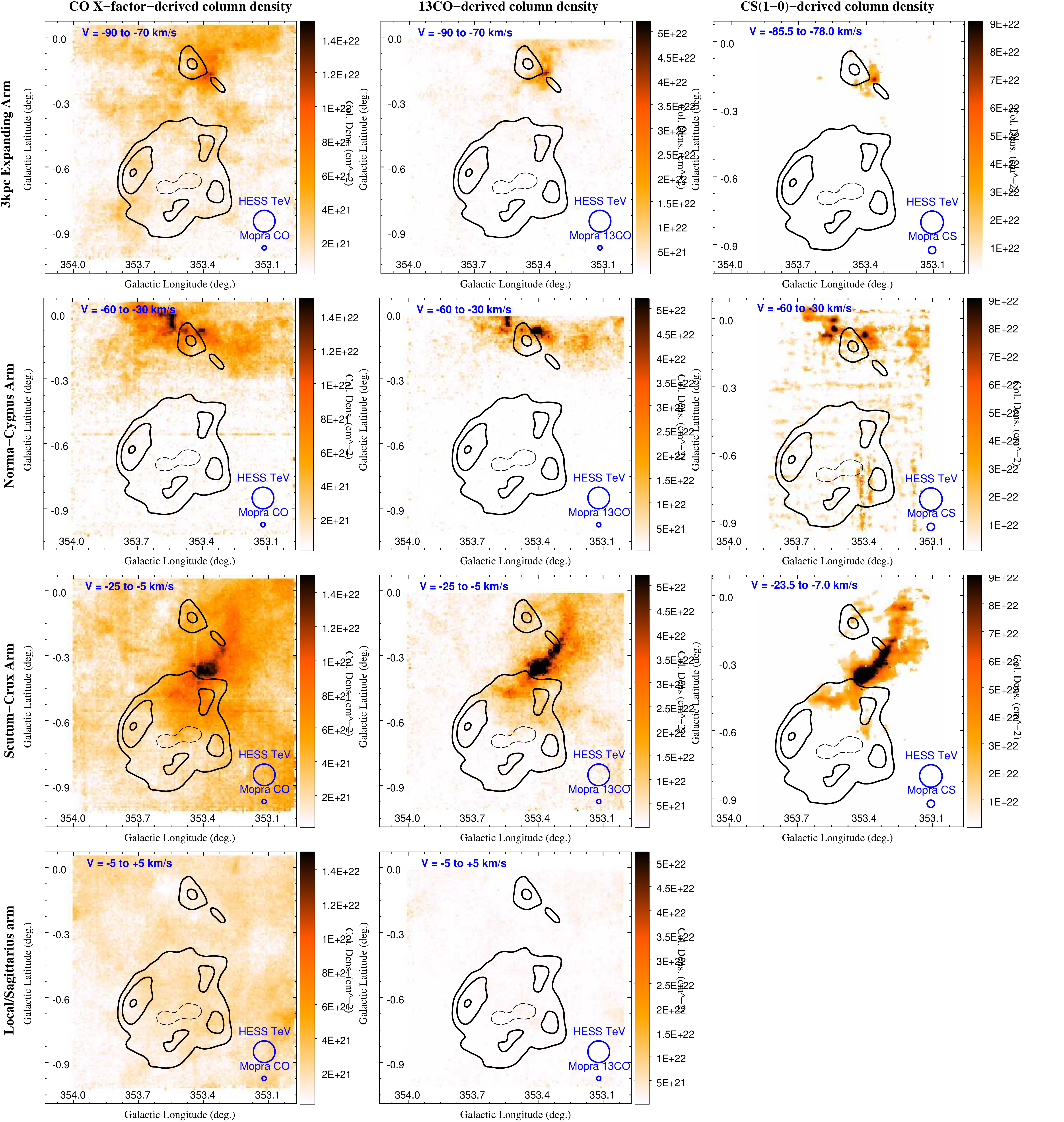}
\caption{The H$_2$ Column density calculated via 3 different methods (CO X-factor, $^{13}$CO analysis, CS analysis, see Sections \ref{sec:Xfactor}, \ref{sec:13CO} and \ref{sec:CS}) towards $\Jsto$ and $\Jstn$ for velocity ranges corresponding to four Galactic arms, the 3kpc-Expanding ($-$90 to $-$70\,$\kms$, top), Norma-Cygnus ($-$60 to $-$30\,$\kms$), Scutum-Crux ($-$25 to $-$5\,$\kms$) and Sagittarius ($-$5 to $+$5\,$\kms$, bottom) arms. HESS 4, 6 and 8$\sigma$ $>$1\,TeV gamma-ray significance contours (thick black) are overlaid \citep{Abramowski:2011}. Thin-dashed contours indicate a central 4$\sigma$ void. Precise velocity range and instrument beam FWHM of each image are indicated. 
\label{fig:ColDens}}
\end{figure*}

This paper utilises three techniques to calculate column density from CO, $^{13}$CO and CS data cubes. Results from these three techniques are displayed in Figure\,\ref{fig:ColDens}. In addition an investigation of $\Jsto$, we are taking steps towards a comprehensive publicly-available molecular density map of the inner Galactic Plane.

In sections\,\ref{sec:13CO} and \ref{sec:CS}, methods that allow the probing of dense molecular gas are outlined. Values from these methods are exploited in Section\,\ref{sec:MassScut} to examine masses of potential CR target material. Firstly, in Section\,\ref{sec:Xfactor}, we describe the use of X-factors, which form a consistent prescription for an unbiased examination of mass components towards sight-lines without accounting specifically for chemically unique environments (i.e. without large star-formation cores).

\subsection{X-factor Analysis}\label{sec:Xfactor}
In the X-factor analysis \citep{Lang:1980}, emission is converted into column density via the equation $N=X\int I_v dv$~~[cm$^{-2}$(K\,$\kms$)$^{-1}$], where $I_v dv$ is the velocity-integrated emission, $X$ is the X-factor and $N$ is the atomic or molecular column density for HI or CO(1-0) emission, respectively. This method effectively utilises a Galactic average for abundance, optical depth and temperature, assumptions that naturally arise from the X-factor derivation that infers a conversion factor from large-scale observations. 

Molecular column densities are derived using the CO-to-H$_2$ mass conversion factor $1.5 \times 10^{20}\mathrm{cm}^{-2}.\mathrm{(K.km.s}^{-1})^{-1}$ from \citet{Strong:2004}, while atomic H densities are derived using the HI brightness temperature to column density conversion factor of $1.8 \times 10^{18}\,\mathrm{cm}^{-2}.\mathrm{(K.km.s}^{-1})^{-1}$ from \citet{Dickey:1990}.

\subsection{$^{13}$CO Analysis}\label{sec:13CO}
$^{13}$CO molecules are $\sim$69$\times$ less abundant than CO, thus $^{13}$CO(1-0) emission suffers less attenuation than CO(1-0). $^{13}$CO emission was analysed via the prescription in \citet{Wilson:2013}. In this procedure, CO(1-0) emission was used to create an excitation temperature map, which was then used to derive the $^{13}$CO(1-0) optical depth. An implicit assumption is that $^{13}$CO and $^{12}$CO emission are both detected from the same region. The $^{13}$CO column density, calculated from the $^{13}$CO optical depth and CO excitation temperature, was then converted to H$_2$ column density assuming a $^{13}$CO abundance of 2.8$\times$10$^{-6}$ \citep[or more commonly quoted as the H$_2$ abundance relative to $^{13}$CO of 3.6$\times$10$^{5}$,][]{Frerking:1982,Bachiller:1986,Cernicharo:1984}.

\subsection{CS Analysis}\label{sec:CS}
For velocity ranges with strong detections of CS(1-0) emission, dense molecular gas could be probed. Column densities were generated from CS(1-0) maps by applying Eq.\,9 of \citet{Goldsmith:1999} to convert CS(1-0) integrated emission into CS(J=1) column density,
\begin{equation}
N_{CS(J=1)}=\frac{8k\pi \nu_{1-0} ^2}{A_{1-0}hc^3}\frac{\Delta\Omega_a}{\Delta\Omega_s}\int{T_{mb}dv}\frac{\tau}{1-e^{-\tau}}
\end{equation}\label{equ:Goldsmith} where CS is described as a linear rotor where $\nu_{1-0}$, $A_{1-0}$, ${\Delta\Omega_a}/{\Delta\Omega_s}$, $T_{mb}dv$, $\tau$ are the CS(1-0) frequency, Einstein coefficient for spontaneous emission, beam filling factor, integrated intensity and CS(1-0) optical depth, respectively. In the maps presented in this paper, the beam-filling factor is assumed to be equal to 1. 
CS(1-0) emission was assumed to be optically thin unless a detection of C$^{34}$S(1-0) emission was recorded in the same location, in which case optical depth was estimated \citep[see e.g. appendix of][]{Maxted:2012} assuming a CS/C$^{34}$S ratio equivalent to the solar system elemental abundance of 22.5. We note that there is some indication of a deviation from this value towards the Galactic centre \citep{Chin:1996}, but this is not considered since C$^{34}$S is only detected within the nearest Galactic arm in this study (see section\,\ref{sec:ScutArm}). 
With the assumption of Local Thermodynamic Equilibrium (LTE) at a temperature of 20\,K \citep[consistent with dust temperatures towards the region, ][]{Schlafly:2011}\footnote{\hbox{http://irsa.ipac.caltech.edu/applications/DUST/}}, the CS(J=1) column density was converted into total CS column density, $N_{CS}$, (CS column density $\sim$5$\times$ CS(J=1) column density). A 50\% error in the assumed temperature of 20\,K would result in a $_{-30}^{+15}$\% systematic error in the column density using this method.

Assuming an abundance of CS with respect to molecular hydrogen, [CS]/[H$_2$]$\sim$10$^{-9}$, a hydrogen column density map could be generated, allowing the estimation of H$_2$ mass. This assumption is generally representative of available data and variation by more than an order of magnitude is uncommon \citep{Frerking:1980,Linke:1980,Shirley:2003,Tafalla:2004}, but possible \citep[e.g.][]{Beuther:2002,Shirley:2003}.

\subsection{Mass estimations}
By applying each of the three analyses outlined above (X-factor, $^{13}$CO, CS), three H$_2$ column density maps were created for each velocity range under scrutiny (Figure\,\ref{fig:ColDens}). These were treated as independent measurements, although the $^{13}$CO analysis is semi-dependent, due to a CO excitation temperature estimation. 

Mass values are quoted where relevant in this paper. H$_2$ column density maps were converted to maps of mass/pixel by $M_{H_2} = 2 m_H A_{pix} N_{H_2} $, where $m_H$ is the mass of a H atom, $A_{pix}$ is the area of a fits file pixel and $N_{H_2}$ is the H$_2$ column density, allowing regions to be simply summed using region statistics functions of a fits-file viewer\footnote{http://ds9.si.edu}.We note that we rejected false emission components that resulted from systematic baseline effects in CS-derived mass maps by subtracting affected regions.


\section{Molecular Gas Towards $\Jsto$ and $\Jstn$}
Gas components at line of sight velocities ranging between $-$100 and 20\,kms$^{-1}$ towards the $\Jsto$ region are presented in Figure\,\ref{fig:PVplot}, which displays CO(1-0) emission corresponding to the local/Sagittarius ($\sim$0\,kms$^{-1}$), Scutum-Crux ($\sim -$17\,kms$^{-1}$), Norma-Cygnus ($\sim -$50\,kms$^{-1}$) and 3kpc-Expanding/Perseus ($\sim -$80\,kms$^{-1}$) arms, consistent with spiral arm modelling by \citet{Vallee:2014_AJ,Vallee:2016} and previous $\Jsto$ gas analyses \citep[e.g.][]{Fukuda:2014}. Such a clear distinction between Galactic arms in Mopra CO(1-0) data so close (6-7$^{\circ}$) to the Galactic centre is notable.

The velocity components seen in Figure\,\ref{fig:PVplot} are individually mapped in Figure\,\ref{fig:mom0s_1} in CO(1-0), $^{13}$CO(1-0) and CS(1-0). 
CO(1-0) traces molecular gas, while $^{13}$CO(1-0) can probe regions where CO(1-0) has become optically thick. CS(1-0) has a similar advantage in CO-optically-thick regions, but also has a critical density for excitation ($\sim$10$^4$\,cm$^{-3}$) that is an order of magnitude larger than CO. It follows that CS only highlights the densest molecular regions. The characteristics of CO, $^{13}$CO and CS are reflected in Figure\,\ref{fig:mom0s_1}, where CO(1-0) emission is most intense and extensive across the field, while $^{13}$CO(1-0) emission is only seen to emanate from regions intense in CO(1-0). CS(1-0) is seen to be even less extensive than $^{13}$CO(1-0), highlighting dense molecular cores in 3 Galactic arms.

Molecular gas in each of the Galactic arms - the Scutum-Crux, the 3kpc-Expanding and the Norma-Cygnus arms are addressed individually in Subsections\,\ref{sec:ScutArm}, \ref{sec:3kpcArm} and \ref{sec:NormaArm}, respectively. In the following subsection (\ref{sec:Xray}) the cumulative column density of all of these arms is examined in detail to derive a kinematic distance to $\Jsto$. The sections after utilise a Scutum-Crux arm association for $\Jsto$ to guide a multi-wavelength investigation of the local environment of $\Jsto$.

\subsection{Column densities towards X-ray emission}\label{sec:Xray}
\begin{figure*}
\includegraphics[width=0.49\textwidth, angle=0]{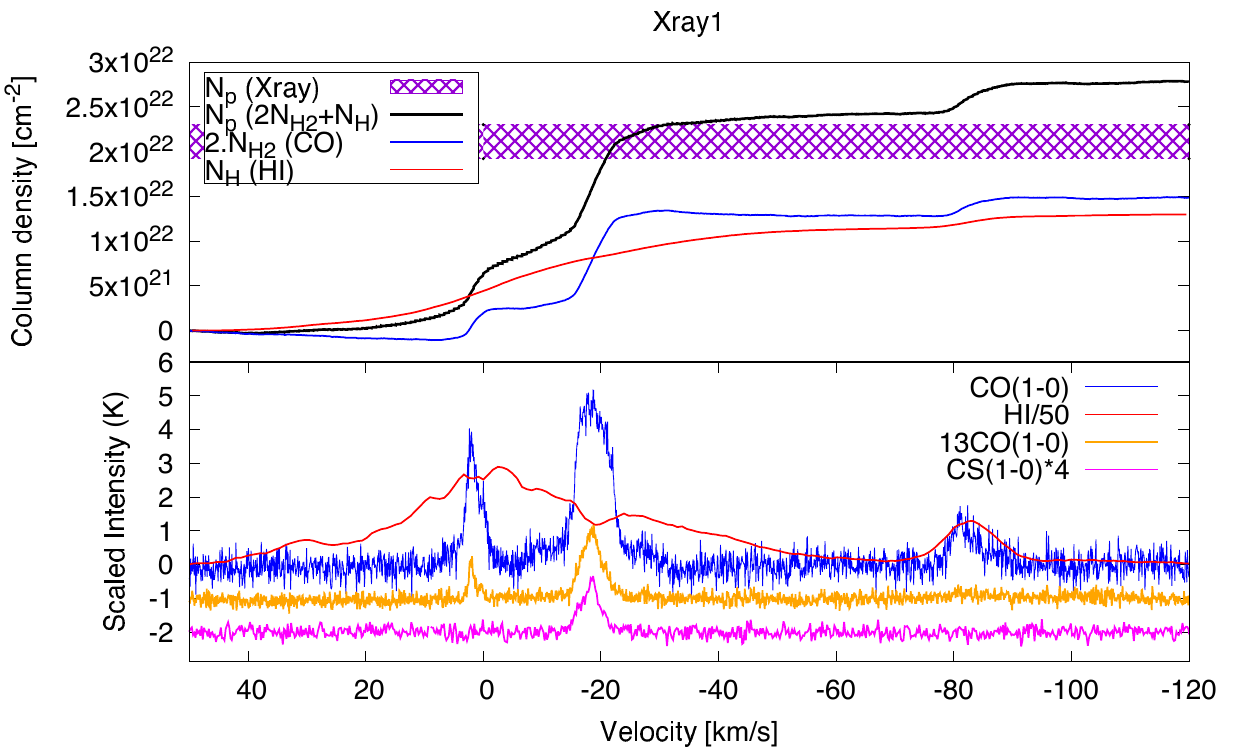}
\includegraphics[width=0.49\textwidth, angle=0]{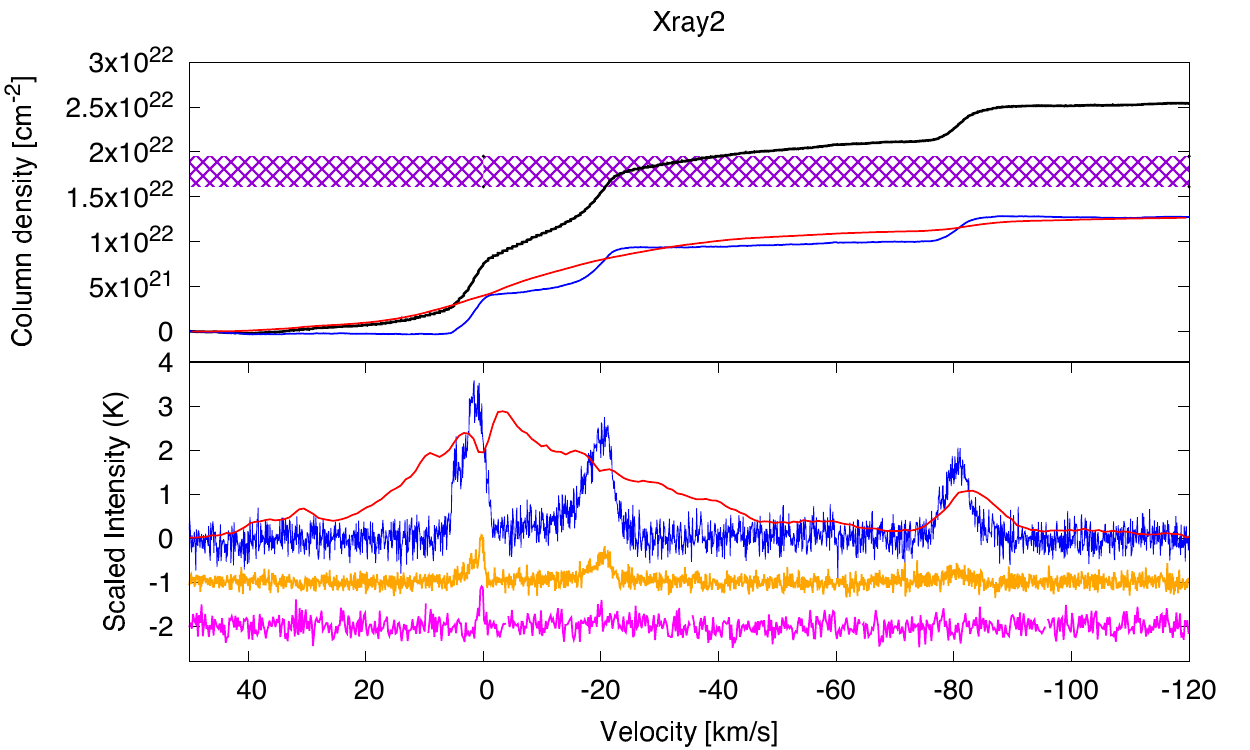}\\
\includegraphics[width=0.49\textwidth, angle=0]{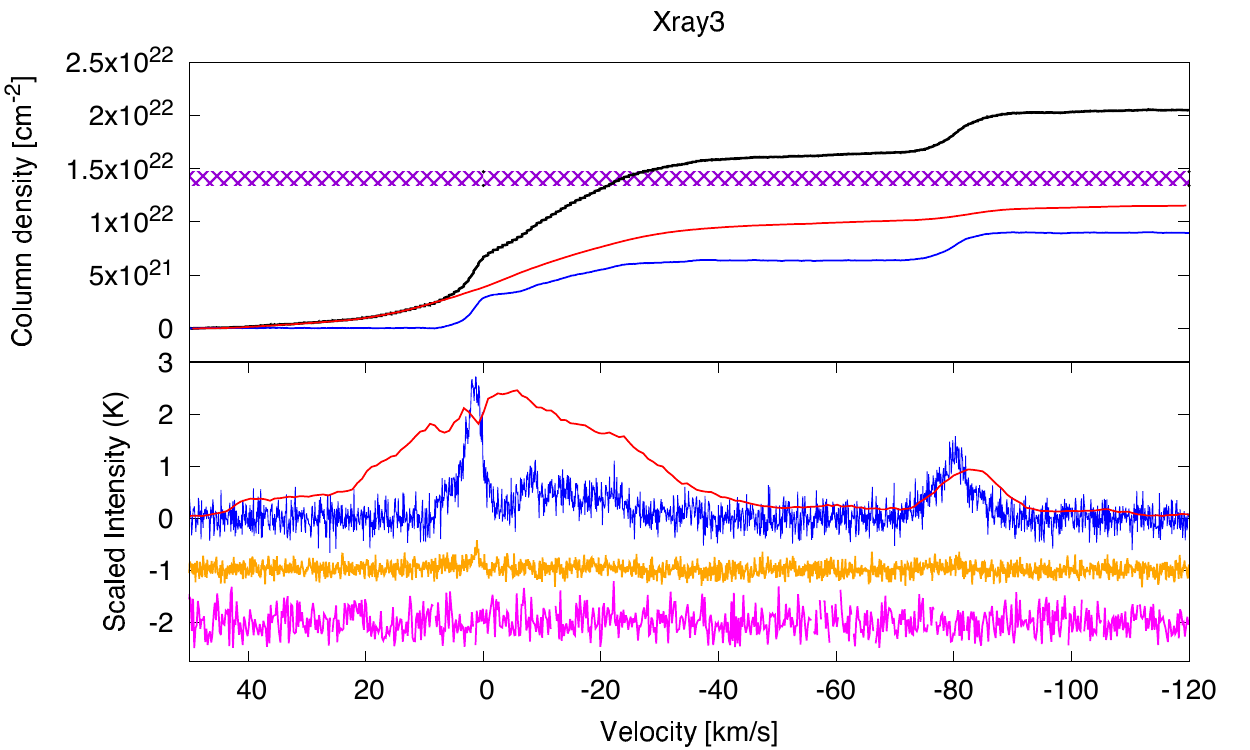}
\includegraphics[width=0.49\textwidth, angle=0]{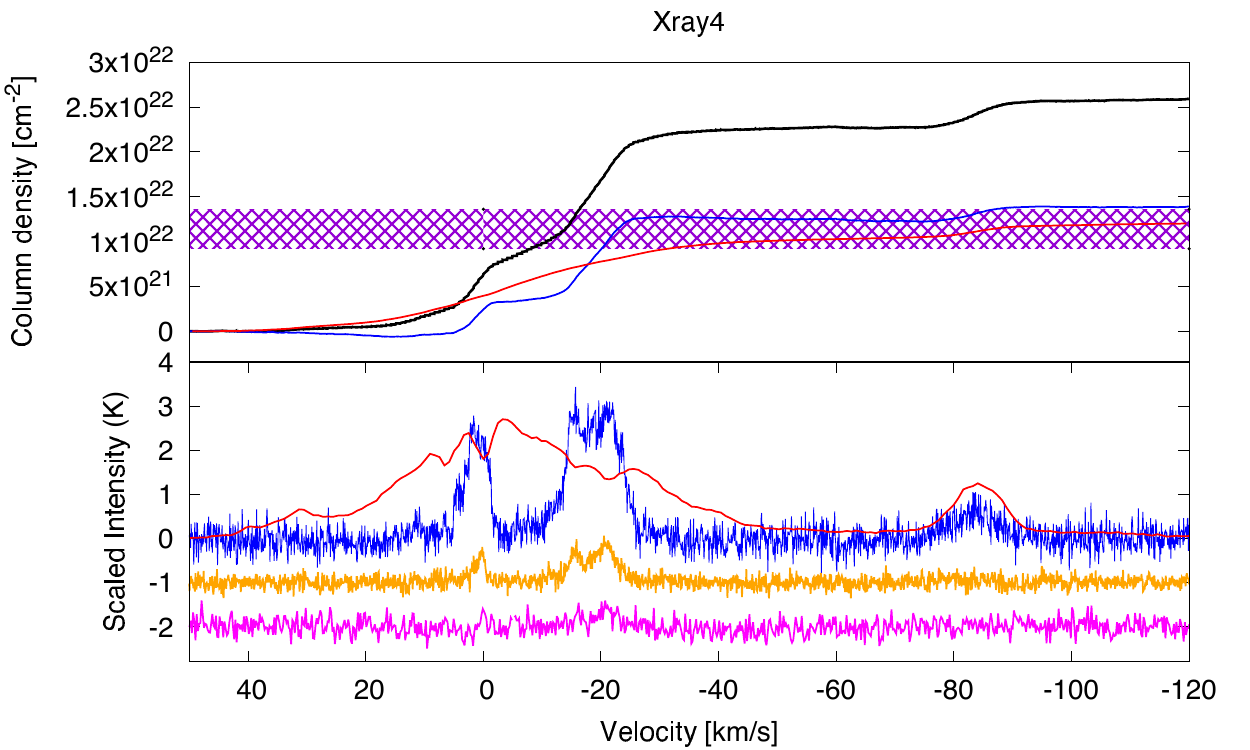}\\
\includegraphics[width=0.49\textwidth, angle=0]{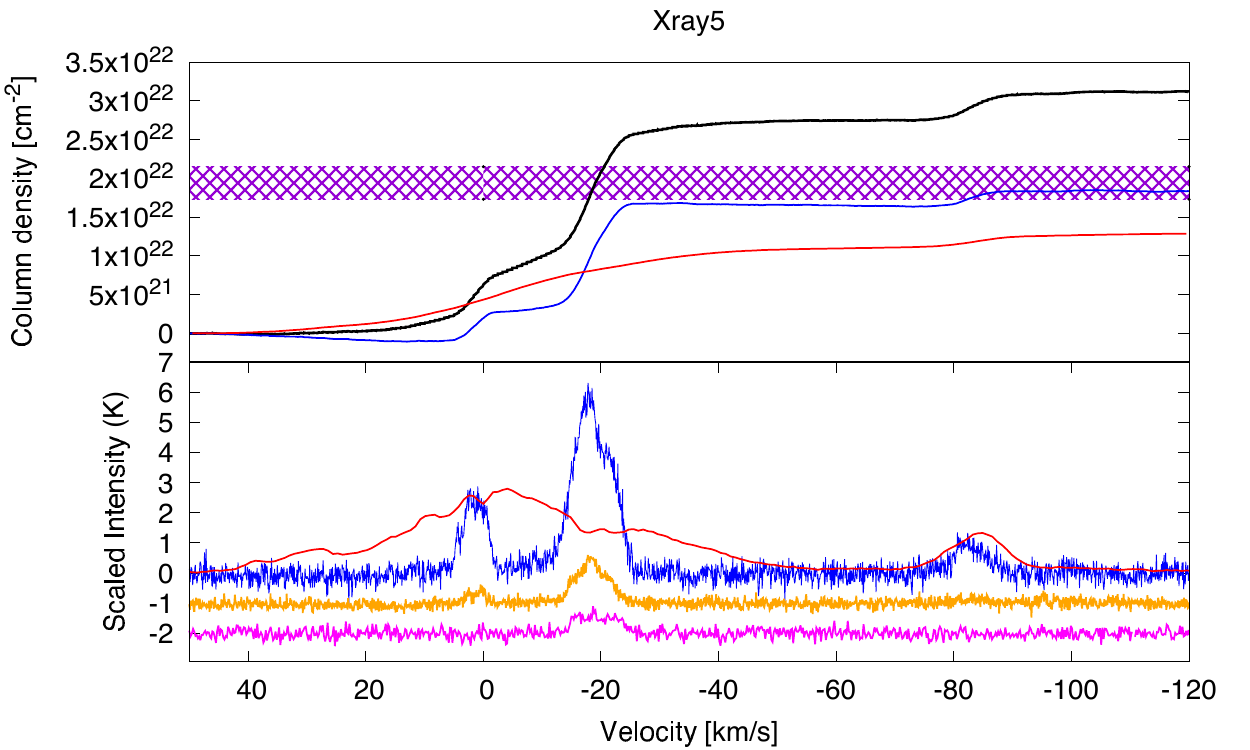}
\includegraphics[width=0.49\textwidth, angle=0]{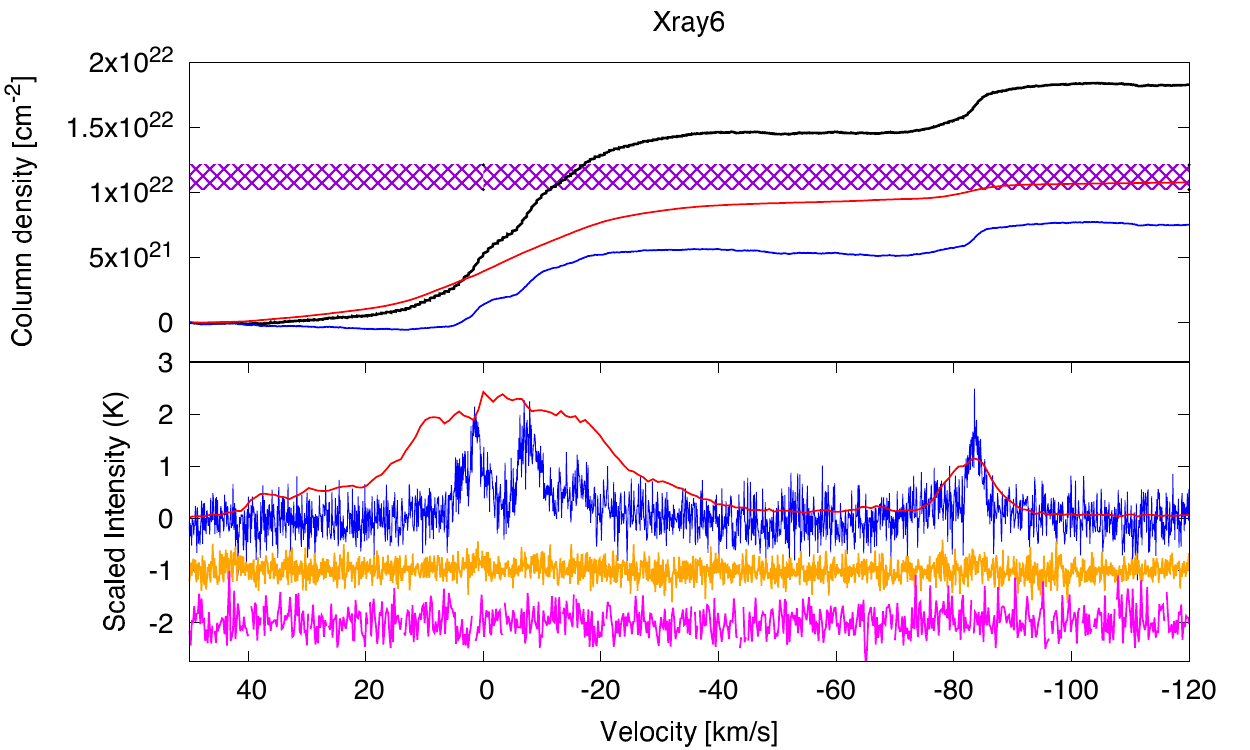}\\
\includegraphics[width=0.49\textwidth, angle=0]{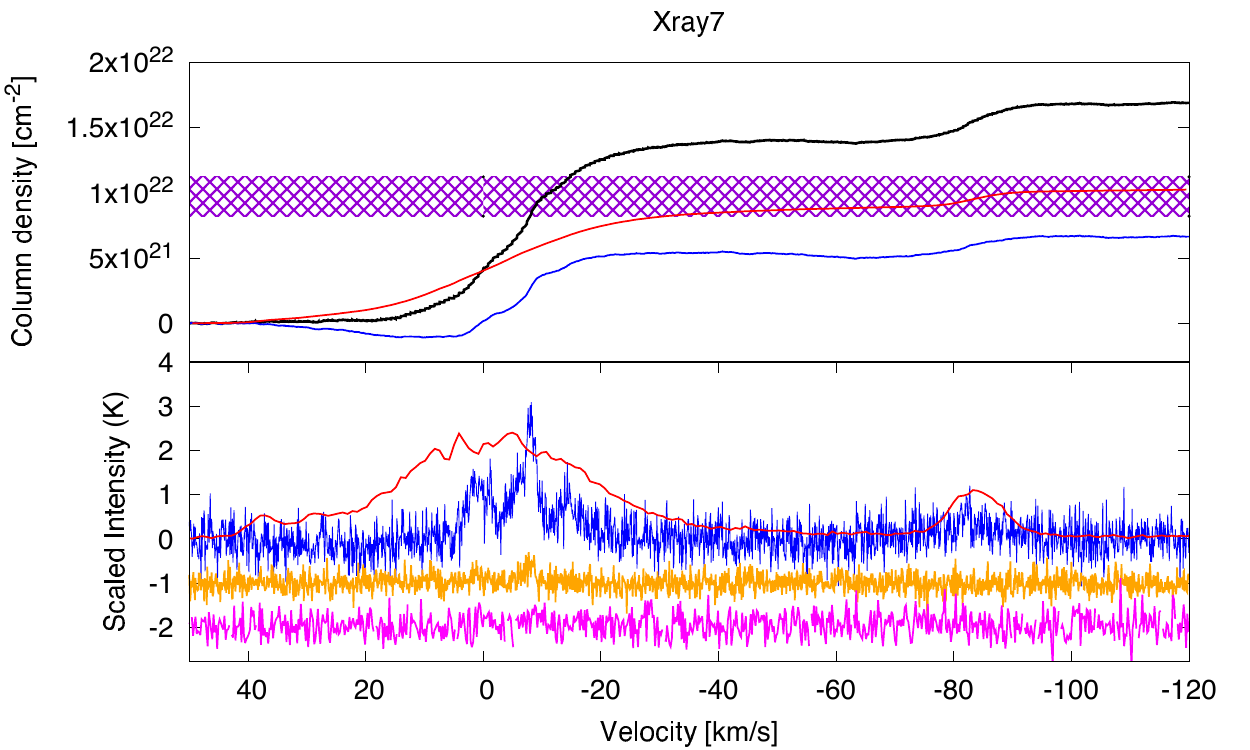}
\includegraphics[width=0.49\textwidth, angle=0]{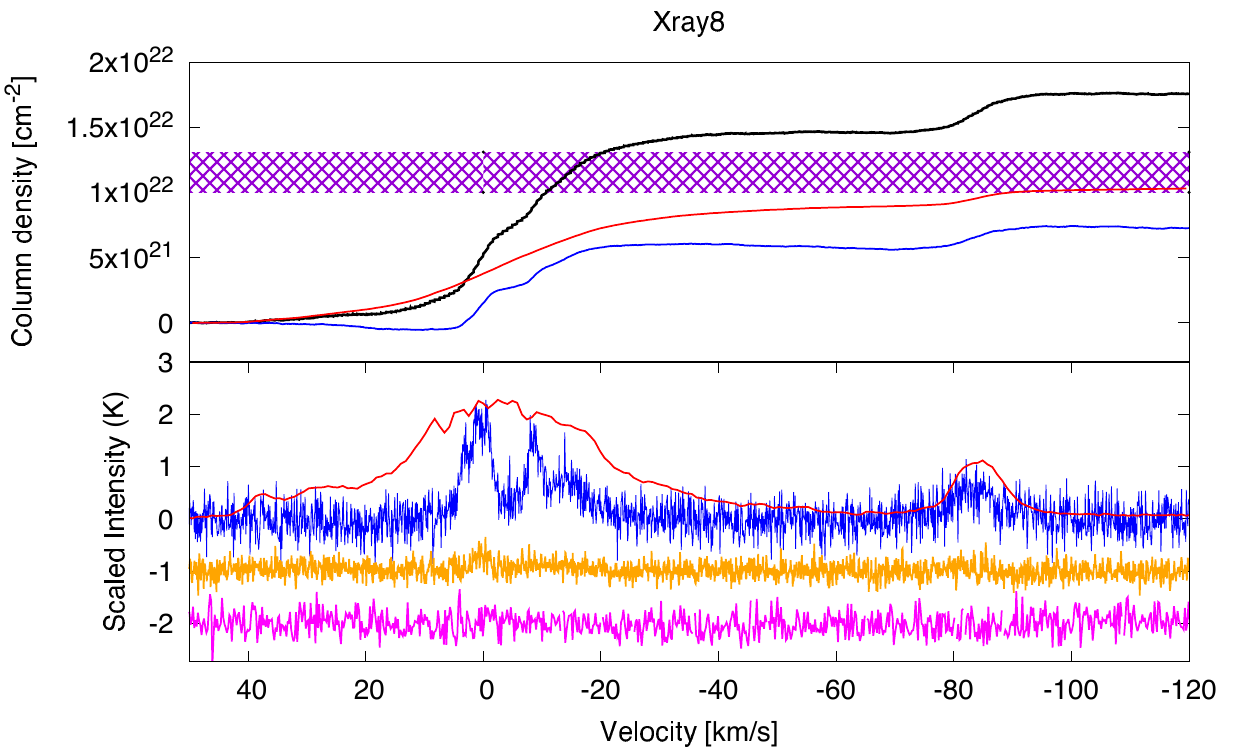}\\
\caption{CO and HI-derived cumulative column density (see Section\,\ref{sec:Xfactor}) plots for regions with $\Jsto$ X-ray emission utilised in \citet{Bamba:2012} to estimate X-ray absorption column densities. For each location, HI, CO(1-0), $^{13}$CO(1-0) and CS(1-0) spectra are displayed in the bottom graphs (red, blue, orange and magenta, respectively), while in the top graphs the 90\%-confidence X-ray absorption column density range is displayed (magenta hatching) with CO and HI-derived column density calculated cumulatively with decreasing velocity (blue and red, respectively).} \label{fig:XraySpecs1}
\end{figure*}

\begin{figure*}
\includegraphics[width=0.49\textwidth, angle=0]{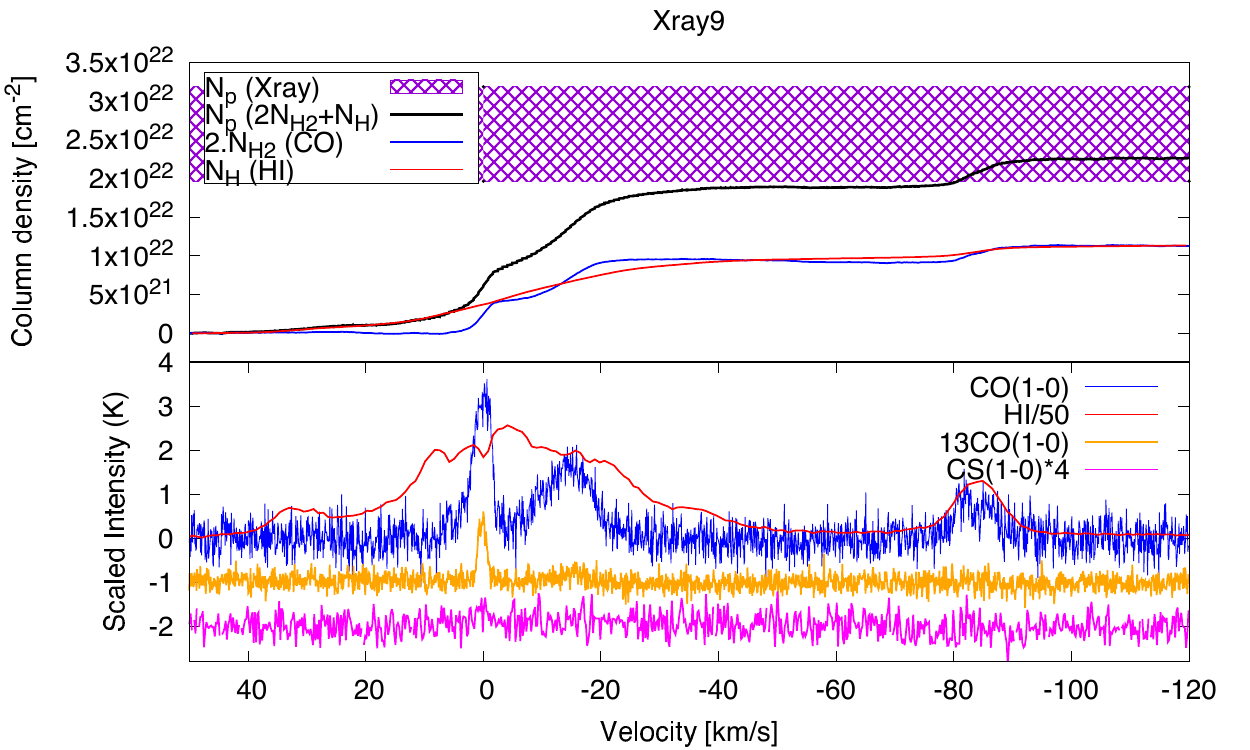}
\includegraphics[width=0.49\textwidth, angle=0]{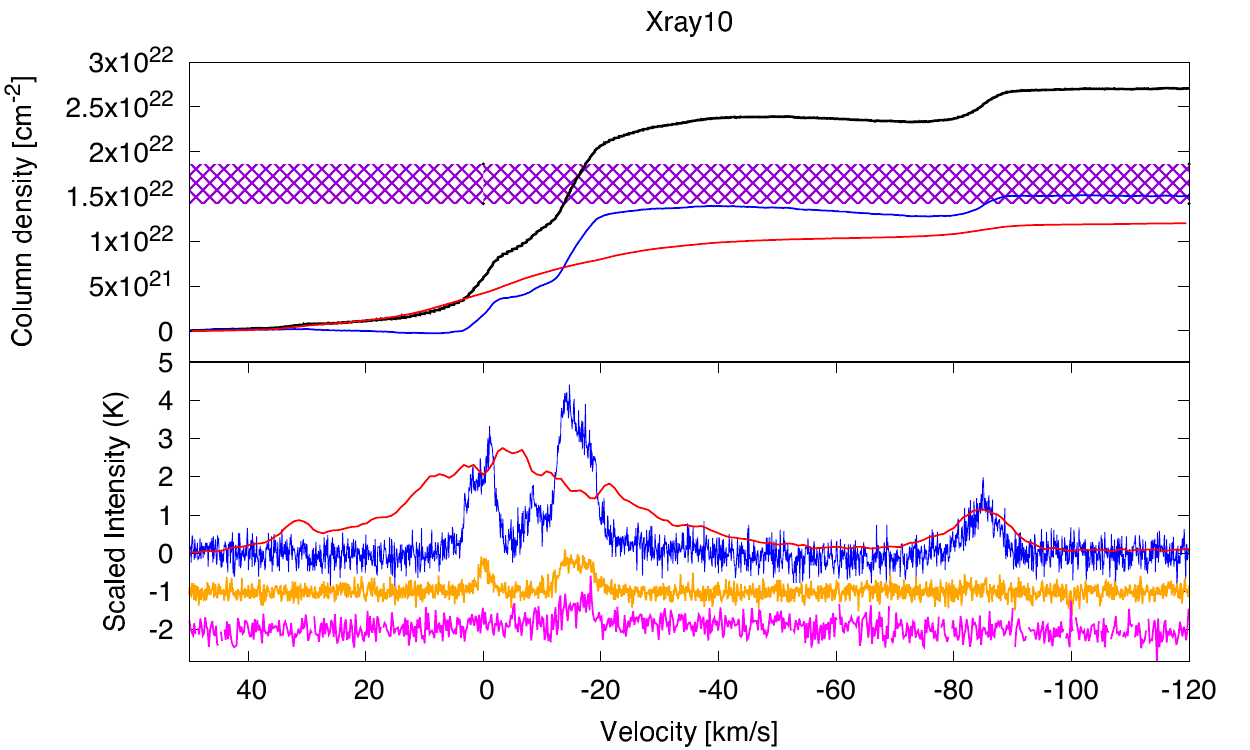}\\
\includegraphics[width=0.49\textwidth, angle=0]{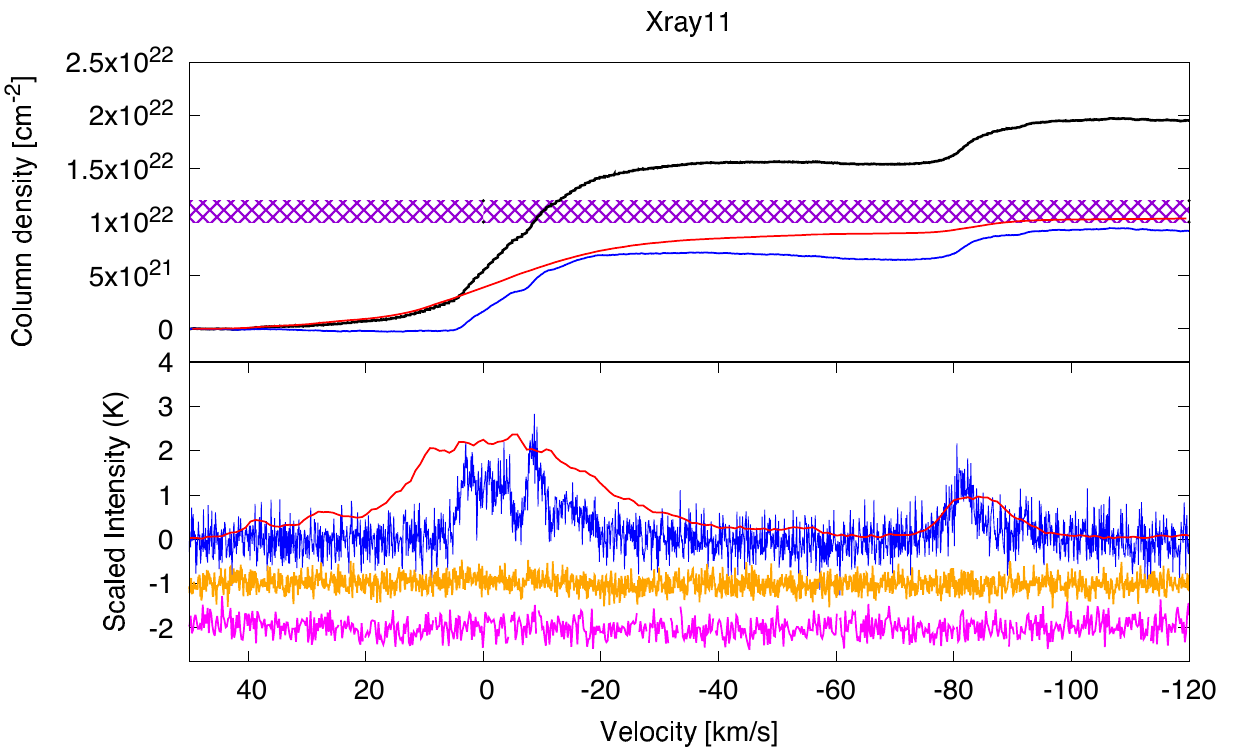}
\includegraphics[width=0.49\textwidth, angle=0]{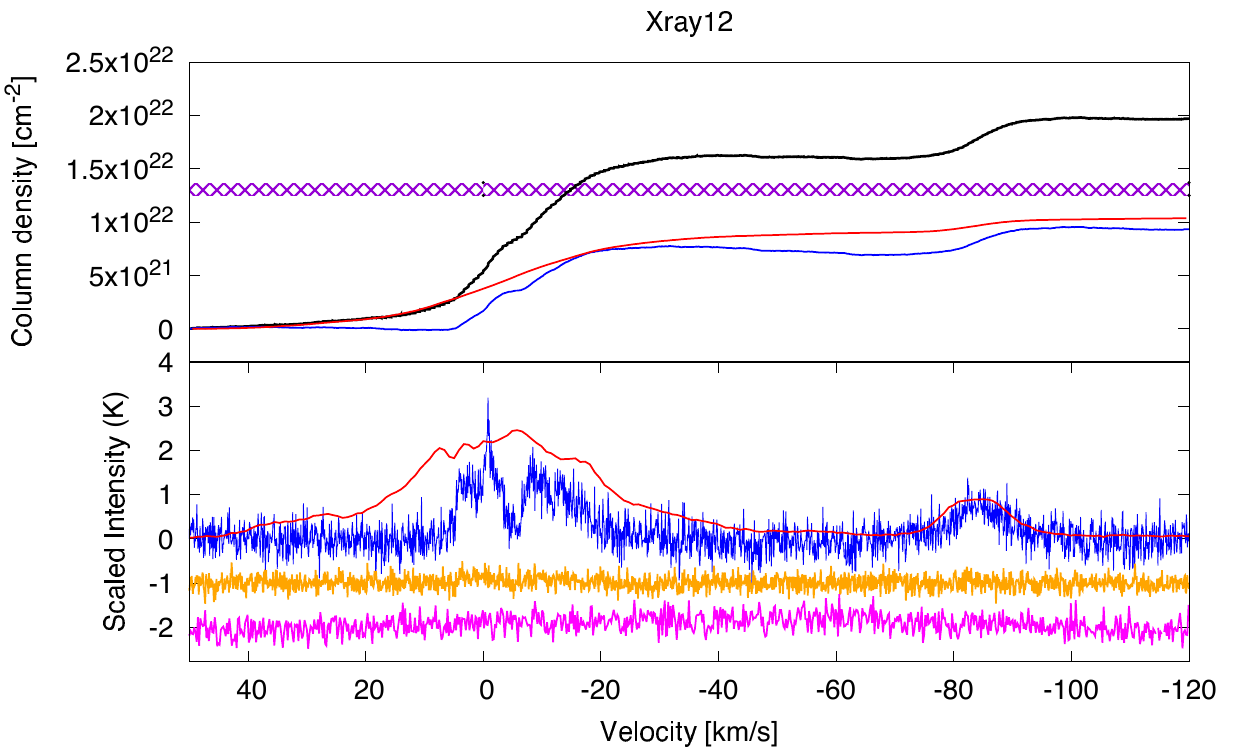}\\
\includegraphics[width=0.49\textwidth, angle=0]{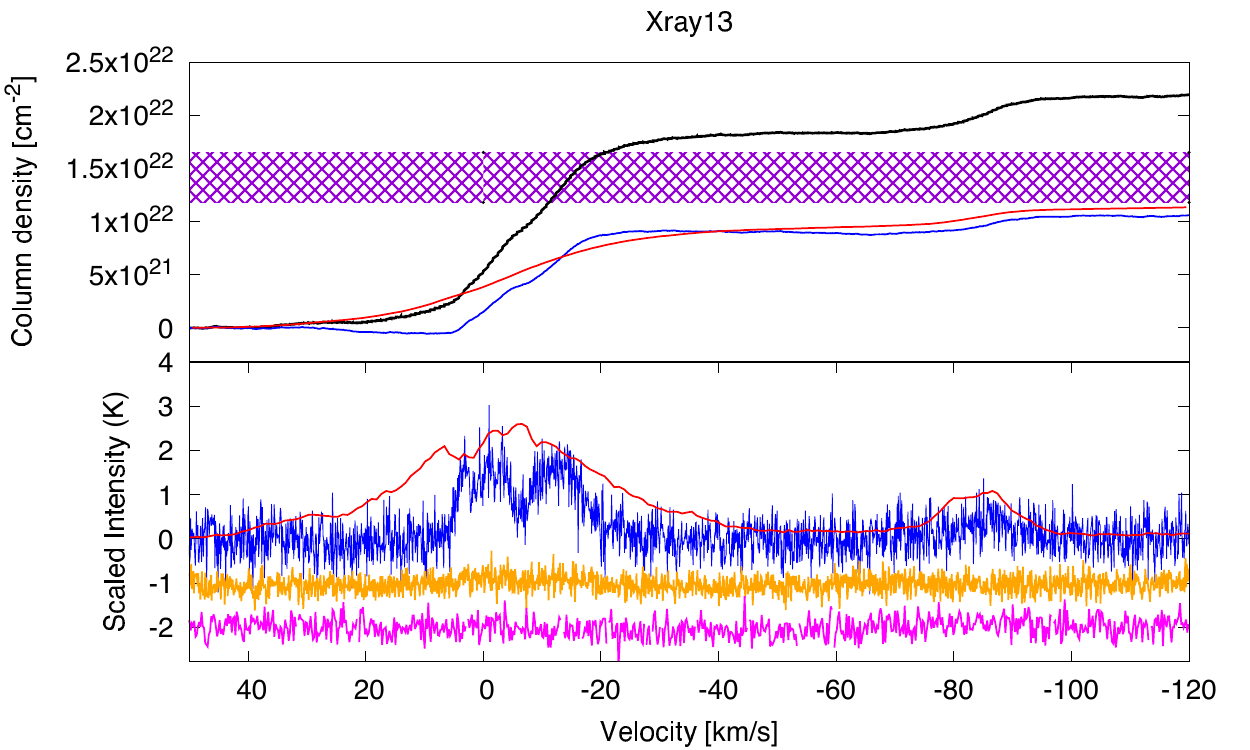}
\includegraphics[width=0.49\textwidth, angle=0]{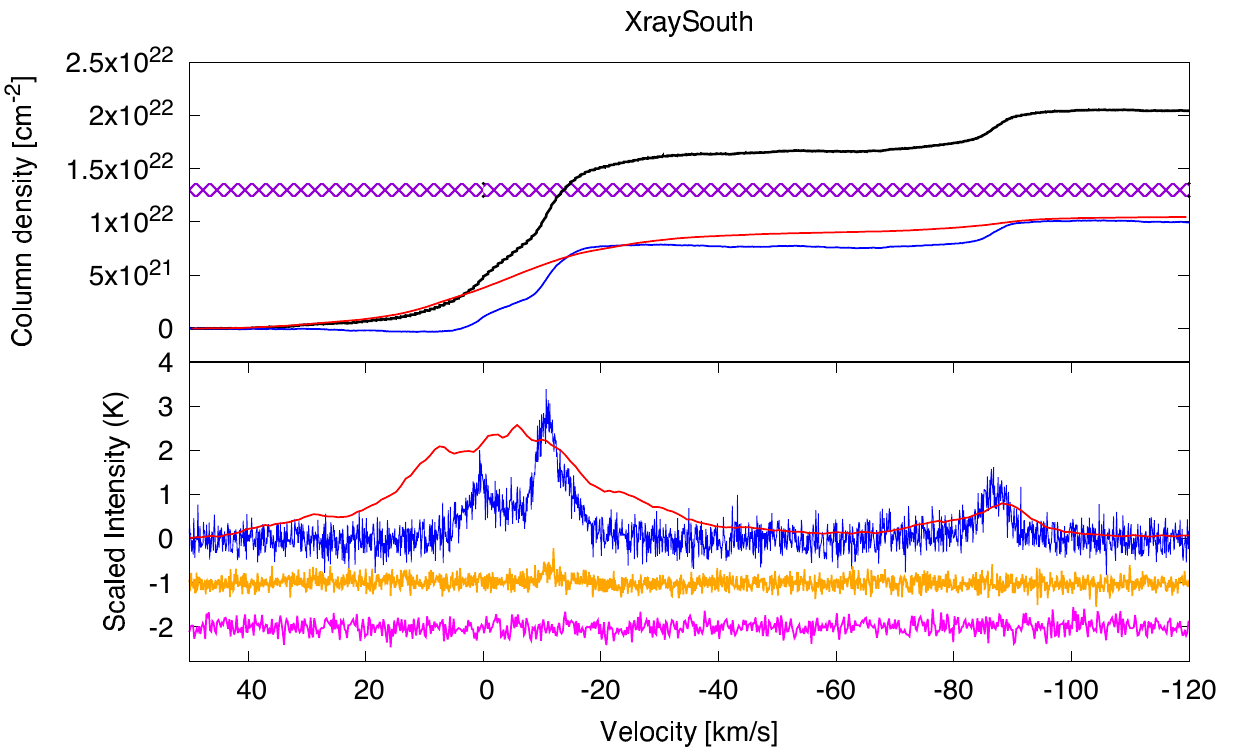}\\
\caption{Same as Figure\,\ref{fig:XraySpecs1}.}\label{fig:XraySpecs2}
\end{figure*}

Figures\,\ref{fig:XraySpecs1} and \ref{fig:XraySpecs2} display graphs of the cumulative column density along the line of sight towards 14 regions with X-ray absorption column densities calculated by \citet{Bamba:2012}. These regions are displayed in Figure\,\ref{fig:1}. X-ray column densities were found through the modelling of photoelectric absorption of non-thermal X-ray photons from a simple absorbed power-law model at 0.5-8\,keV. A comparison of these values with column densities derived from Mopra CO(1-0) emission and SGPS HI emission can help to constrain the kinematic distance to $\Jsto$. We note that the fourteen X-ray emission regions examined have scales of $\sim$3-4$^{\prime}$, making the fine angular resolution of Mopra CO and SGPS HI data ($\sim$35$^{\prime\prime}$ and $\sim$2$^{\prime}$, respectively) a precise comparative tool at arcmintute scales. 
When calculating column densities (see Section\,\ref{sec:Xfactor}) it was assumed that all the spectral emission towards $\Jsto$ is near-side, and that emission with velocity centroids decreasing from $\sim +$10 to $\sim -$100\,$\kms$ correspond to increasing distance from $\sim$0.5 to $\sim$6\,kpc, consistent with previous analyses \citep{Tian:2008,Tian:2010,Fukuda:2014}. This assumption is also supported by the presence of HI dips corresponding to self-absorption at local/Sagittarius and Scutum-Crux arm velocities ($\sim +$10 to $-$25\,$\kms$). 
We note that if this assumption leads to the inclusion of a component of far-side gas in cumulative column density calculations, the conclusions will still hold in the case that such a component does not greatly exceed $\sim$20\% (see Appendix\,\ref{Sec:app_distance}).

Figures\,\ref{fig:XraySpecs1} and \ref{fig:XraySpecs2} display both X-ray absorption-derived H column densities (hereafter $N_{p}^{X}$), and CO+HI-derived H column densities (hereafter $N_{p}^{CO+H}$). In principle, $N_{p}^{X}$ traces all of the spectral components foreground to the $\Jsto$ X-ray emission, hence the velocity of the point of convergence of the values derived from the two techniques offers a way to uniquely determine the velocity corresponding to $\Jsto$ (hence kinematic distance).

\begin{figure}
\includegraphics[width=0.51\textwidth, angle=0]{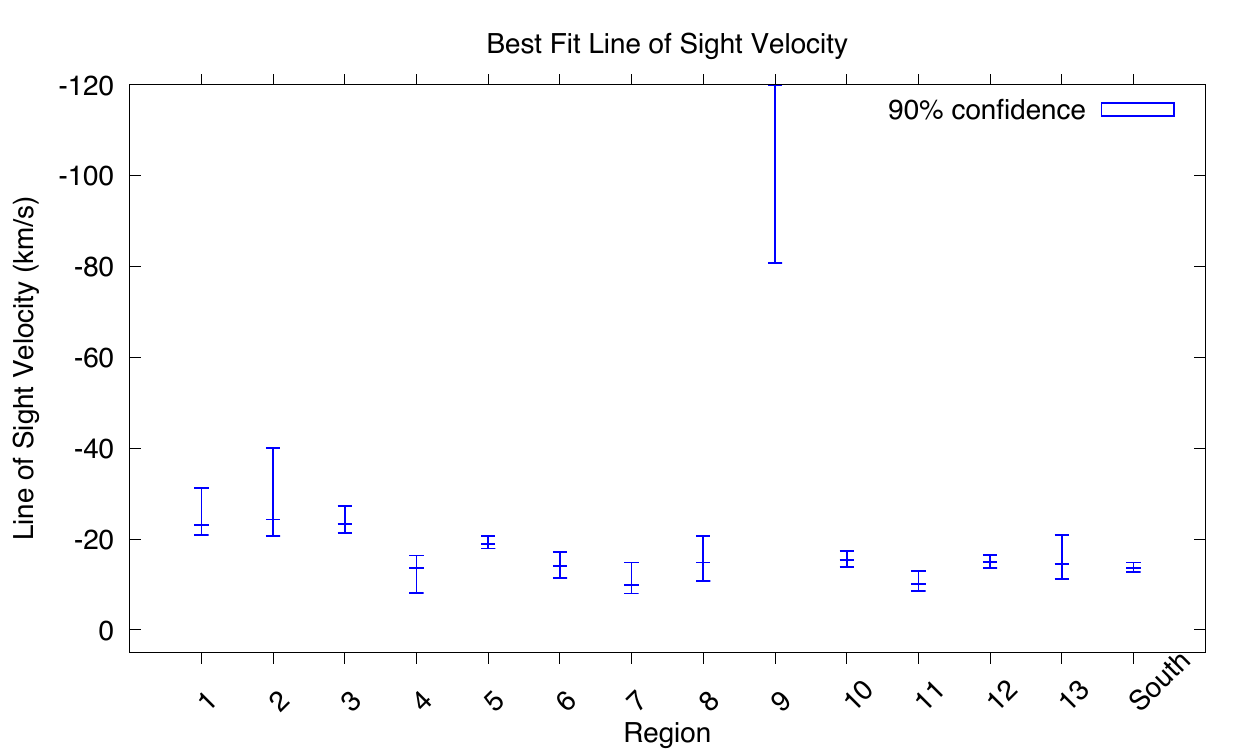}\\
\caption{The velocity of the intercepts of X-ray absorption column densities and corresponding spectrally-derived column densities for 14 regions towards the $\Jsto$ X-ray emission. Blue error bars indicate the 90\% confidence limits of the data. The inverse variance-weighted mean of the data is $\vlsr = -$15.2$\pm$0.4\,$\kms$. 
\label{fig:BoxAndWhiskers}}
\end{figure}

Towards all but one region, $N_{p}^{X}$ converges with $N_{p}^{CO+H}$ at typical values between 1.0 and 2.5$\times$10$^{22}$\,$\cmsqr$ at velocities between $\sim -$5 and $-$25\,$\kms$, indicating that $\Jsto$ is probably associated with this velocity component. Figure\,\ref{fig:BoxAndWhiskers} summarises the results of the comparison of $N_{p}^{X}$ and $N_{p}^{CO+H}$ for all of the regions scrutinised. The weighted arithmetic mean line of sight velocity is $\sim -$15\,$\kms$, consistent with the velocity of the Scutum-Crux arm at 3.2\,kpc. 

We note that one out of the 14 tested positions (`Xray9' in Figure\,\ref{fig:XraySpecs2}) would favour an association with the gas component at $-$80\,$\kms$ (3\,kpc Expanding arm) if the result was taken in isolation, however this value has the largest 90\% certainty interval ($\sim$1.21$\times 10^{22}$\,$\cmsqr$), hence it does not significantly contribute to the weighted arithmetic mean velocity (of $\sim -$15\,$\kms$). We further note, that recent work by \citet{Doroshenko:2017} benefit from better statistics towards `Xray9' leading the authors to revise the value of absorption column density down to $\sim$1.6$\times 10^{22}$\,$\cmsqr$, consistent with the 3.2\,kpc kinematic distance solution. Other X-ray absorption column density values towards $\Jsto$ are consistent with the corresponding \citeauthor{Doroshenko:2017} values. 

We explore the effect of systematic errors in Appendix\,\ref{Sec:app_distance} and conclude that a Scutum-Crux arm association still holds even if sources of uncertainty up to $\sim$20\% exist. Larger unaccounted-for systematic errors may affect the derived kinematic distance, but this depends on the direction of the systematic error.

Our technique does not necessarily have sufficient precision to probe the location of $\Jsto$ beyond a simple flagging of an associated arm (Scutum-Crux), because local velocity components and turbulence probably dominate velocities at the scale of the spectral line-width, unlike peak spectral line velocities which more clearly reflect Galactic kinematic motions that correspond to Galactic model-dependent kinematic distance.

\subsection{Gas in the Scutum-Crux Arm}\label{sec:ScutArm}
Our analysis suggests that gas of the Scutum-Crux arm at 3.2\,kpc ($\sim -$15\,$\kms$) is associated with $\Jsto$, and the analysis following finds an association of Scutum-Crux arm gas with infrared dark regions and possibly radio-continuum emission from $\Jsto$. 
\begin{figure*}
\includegraphics[width=1.02\textwidth, angle=0]{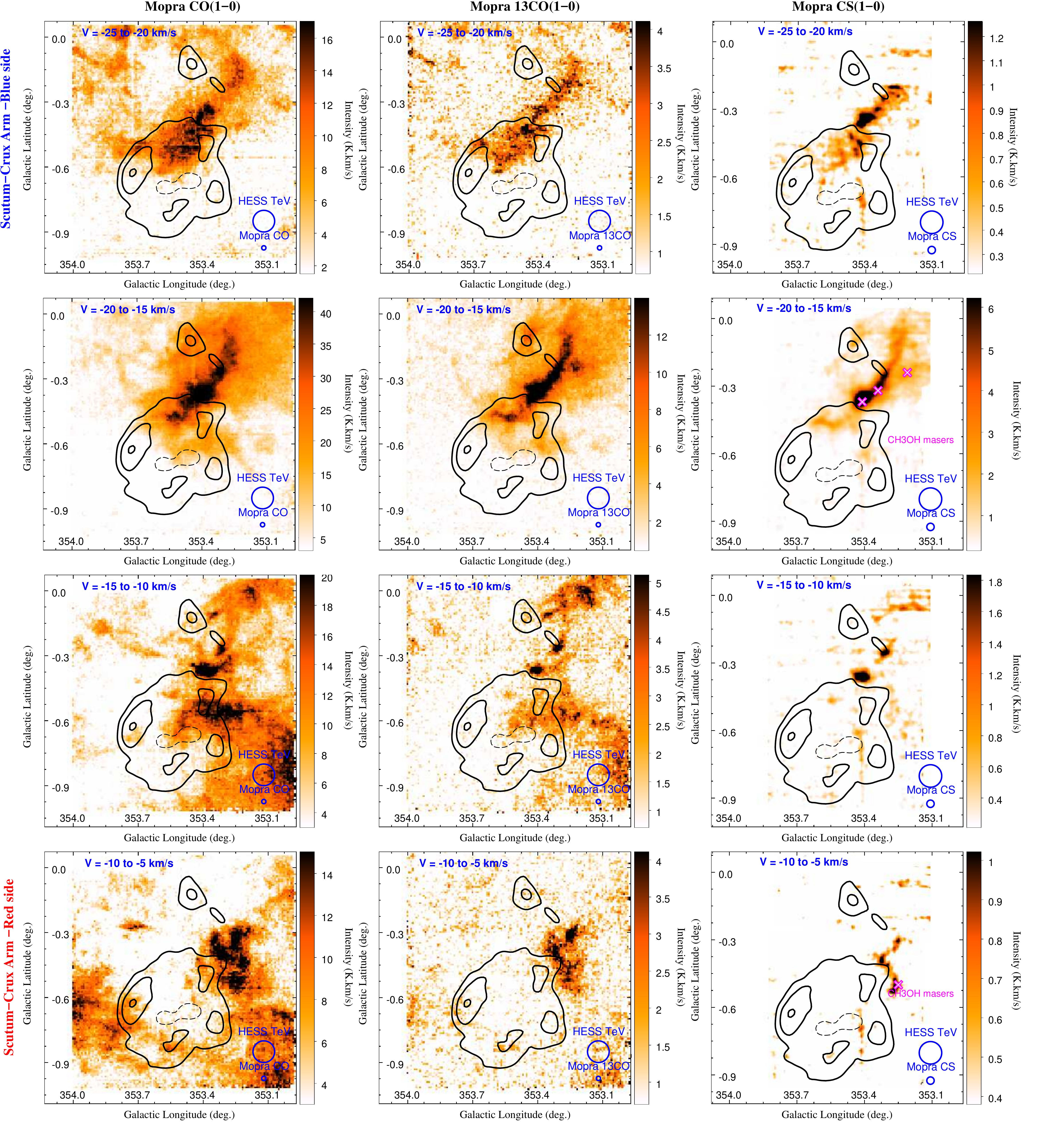}
\caption{Velocity integrated CO(1-0) (left), $^{13}$CO(1-0) (middle) and CS(1-0) (right) emission towards $\Jsto$ and $\Jstn$ for four velocity range slices corresponding to the Scutum-Crux arm ($-$25 to $-$20, $-$20 to $-$15, $-$15 to $-$10, and $-$10 to $-$5\,$\kms$). HESS 4, 6 and 8$\sigma$ $>$1\,TeV gamma-ray excess contours (thick black) are overlaid \citep{Abramowski:2011}. Thin-dashed contours indicate a central 4$\sigma$ void. Precise velocity range and instrument beam FWHM of each image are indicated. The location of CH$_3$OH emission is indicated on CS(1-0) images with corresponding velocity ranges. \label{fig:mom0s_2}}
\end{figure*}

CO(1-0), $^{13}$CO(1-0) and CS(1-0) images of the Scutum-Crux arm, spanning the velocity $\vlsr \sim -$25 to $-$5\,$\kms$, are displayed in Figure\,\ref{fig:mom0s_1} alongside emission from other Galactic arms. We examine the Scutum-Crux range in 5\,$\kms$ slices in Figure\,\ref{fig:mom0s_2}.

\subsubsection{HII region G353.43$-$0.37, $\Jstn$ and the Surrounding Infrared-dark Gas} 
\begin{figure*}
\includegraphics[width=0.99\textwidth, angle=0]{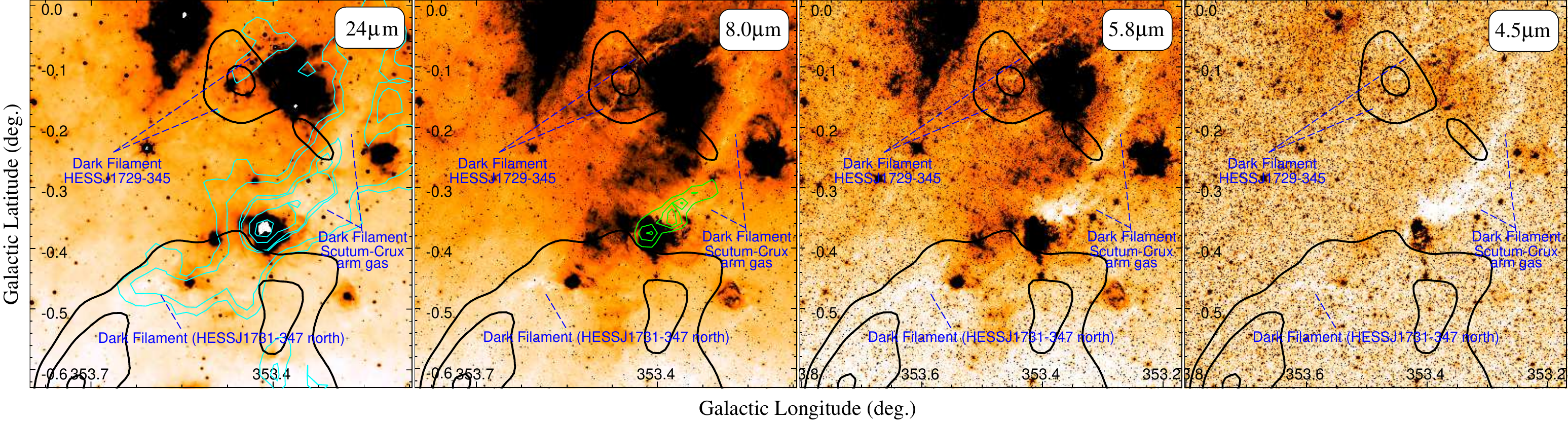}
\caption{24, 8, 5.8 and 4.5\,$\mu$m images of the $\Jsto$/$\Jstn$ region \citep[MIPSGAL and GLIMPSE, ][]{Churchwell:2009,Carey:2009}. Dark and light regions correspond to high and low-intensity regions, respectively. The colour scales have been adjusted to emphasize infrared-dark gas. Cyan contours in the 24\,$\mu$m image correspond to CS(1-0)-derived H$_2$ column density (levels $0.1$-$5\times10^{23}$\,$\cmsqr$, spaced irregularly). Green contours in the 8\,$\mu$m image correspond to CS(1-0) optical depth (levels 0, 1, 2, 3, 4 and 5, where the `0' contour level represents the border between C$^{34}$S(1-0) detections and non-detections). Zoomed-in images of `Dark Filament $\Jstn$' and `Dark Filament $\Jsto$-north' in 8\,$\mu$m infrared emission appear in Figure\,\ref{fig:infraredZoom}.}\label{fig:infrared}
\end{figure*}

Scutum-Crux gas harbours a HII region, G353.42$-$0.37, which has an estimated kinematic distance of $\sim$3.2\,kpc \citep{Tian:2008}. A filament of infrared-dark gas labelled ``Dark Filament Scutum-Crux arm gas'' in 24, 8.0, 5.8 and 4.5\,$\mu$m images of Figure\,\ref{fig:infrared} contains G353.42$-$0.37. A dense clump associated with this HII region can be clearly observed in CO, $^{13}$CO and CS images in Figure\,\ref{fig:mom0s_2}, while the velocity centroid of the encompassing molecular cloud is at $\sim-$17\,$\kms$, as evidenced by CS isotopologues, SiO, CH$_3$OH and HC$_3$N emission (see Appendix\,\ref{sec:app_lines} for spectral line fit parameters). The optical extinction\footnote{calculated at http://irsa.ipac.caltech.edu/applications/DUST/} of this region reaches $A_v \sim$90 \citep[assuming a visual extinction to reddening ratio of 3.1, ][]{Schlafly:2011}, suggesting a column density of $\sim$2$\times$10$^{23}$\,cm$^{-2}$ \citep[using $N_H \sim 2.2\times10^{21} A_v$ \,cm$^{-2}$,][]{Guver:2009}. This infrared-derived value is consistent with the mean column density of $\sim$2.1$\times$10$^{23}$\,cm$^{-2}$ derived from the dense gas tracer, CS(1-0), strongly supporting a connection between the infrared-dark feature and the Scutum-Crux arm gas.

The rich chemistry of the cloud suggests a warm, dense environment that is commonly associated with star-formation or gas irradiated/warmed by stellar activity. The spectral profile of the gas towards the HII region is broad (CS FWHM$\sim$5.8\,$\kms$) and can be seen in 3 velocity slices of Figure\,\ref{fig:mom0s_2} (3 top-most images). 

A filament of infrared-dark gas labelled ``Dark Filament $\Jstn$'' can be seen to cross $\Jstn$ at a $\sim$45$^{\circ}$-angle to the Galactic Plane. The feature, which is seen in all infrared images of Figure\,\ref{fig:infrared} but most prominently in 8 and 5.8\,$\mu$m bands, may correspond to Scutum-Crux arm gas at 3.2\,kpc that is seen most clearly in CS(1-0) emission at $\vlsr\sim -20$ to $-$15\,$\kms$ in Figure \ref{fig:mom0s_2}. The infrared opacity-derived column density of $\sim$2.6$\times$10$^{22}$\,cm$^{-2}$ ($A_v\sim$12), is consistent with the CS(1-0)-derived value of $\sim$2.5$\times$10$^{22}$\,cm$^{-2}$. 
With the detection of this dense gas component towards $\Jstn$, weight is added to the runaway CR scenario for the origin of $\Jstn$ presented by \citet{Cui:2016}. We note that \citet{deWilt:2017} also find a good correspondence between the $\Jstn$ gamma-ray emission and another dense gas tracer, namely NH$_3$(1,1) emission.

An infrared-dark filament, labelled ``Dark Filament $\Jsto$ north'' in Figure\,\ref{fig:infrared}, over-laps with the north of $\Jsto$. Like the aforementioned infrared-dark features, the CS(1-0)-derived column densities are approximately consistent with the opacity-derived measurements ($\sim$6 and $\sim$4$\times$10$^{22}$\,cm$^{-2}$, respectively). This specific filament is addressed further in the next section, but we note that there is a good correspondence between CS-traced filaments and all infrared-dark filaments discussed in this region, as illustrated in Figure\,\ref{fig:infrared}. Zoomed-in images of both `Dark Filament $\Jstn$'' and `Dark Filament $\Jsto$ north'' are displayed in Appendix\,\ref{sec:app_lines}.

\subsubsection{Radio-continuum Correspondance and a Molecular Void}\label{sec:RadContCor}
\begin{figure*}
\begin{centering}
\includegraphics[width=1.02\textwidth, angle=0]{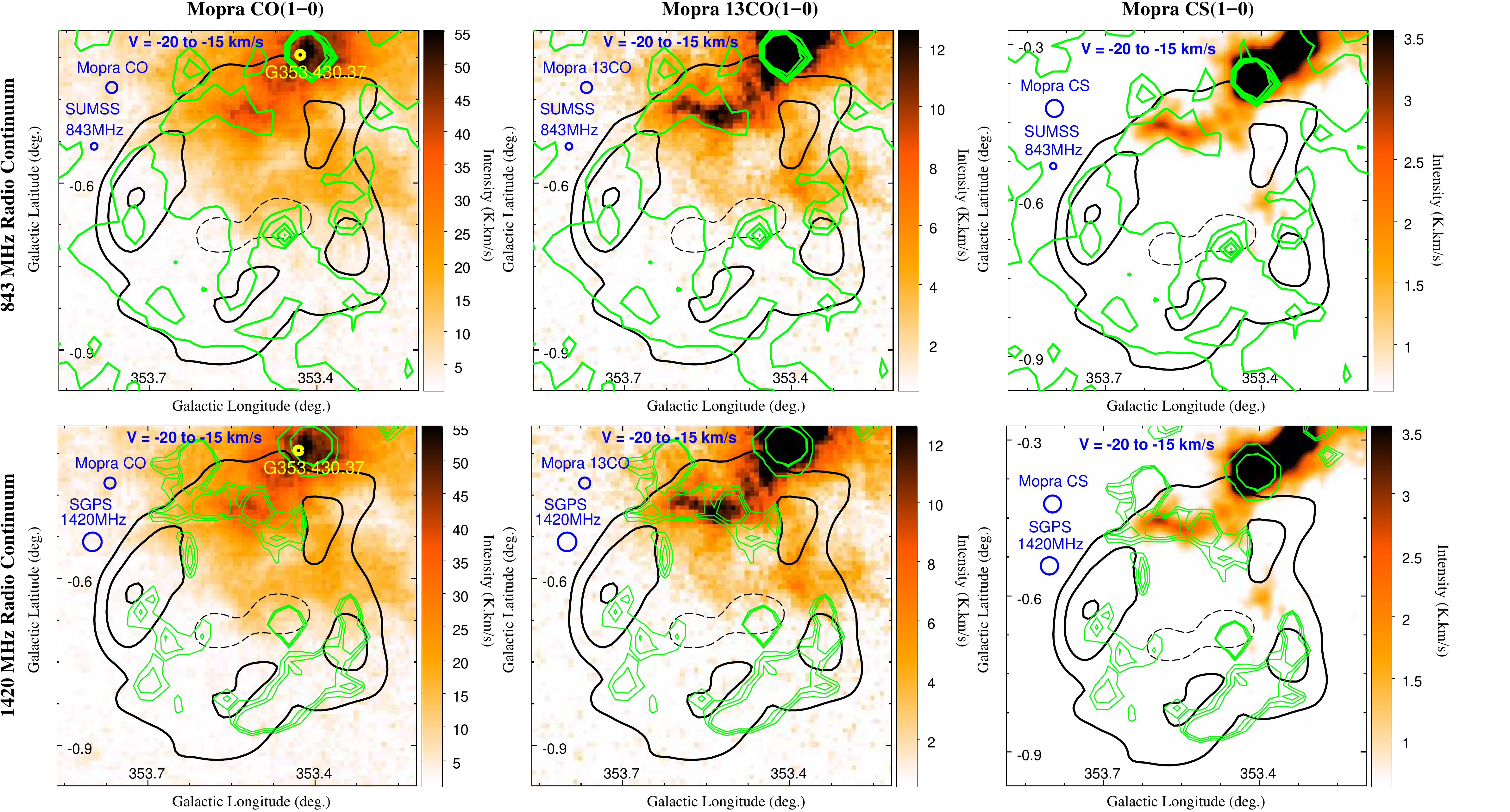}
\caption{CO(1-0), $^{13}$CO(1-0) and CS(1-0) emission integrated between velocities $-$20 and $-$15\,$\kms$. The top images have overlaid 843\,MHz contours \citep{Green:1999}, while the bottom images have overlaid SGPS 1420\,MHz contours \citep{Haverkorn:2006}. HESS 4, 6 and 8$\sigma$ $>$1\,TeV gamma-ray excess contours (thick black) are overlaid \citep{Abramowski:2011}. Thin-dashed contours indicate a central 4$\sigma$ void. Instrument beam FWHM of each image are indicated. In the left-most CO(1-0) image the location of HII region G353.43$-$0.37 is indicated.\label{fig:RadioCont}}
\end{centering}
\end{figure*}

\citet{Tian:2008} identified SNR G353.6$-$0.7, a non-thermal radio continuum structure at 843 and 1420\,MHz that is likely to be associated with $\Jsto$ (see Section\,\ref{sec:intro}). The northern rim of the G353.6$-$0.7/$\Jsto$ radio continuum has a good correspondence with a dense molecular clump traced by $^{13}$CO(1-0) and CS(1-0) at [$l,b$]$\sim$[353.6$^{\circ},-$0.45$^{\circ}$] in Figure\,\ref{fig:RadioCont}. The correspondence is suggestive of a shock-compression at a kinematic distance of $\sim$3.2\,kpc. We further note that this dense gas component, which corresponds to the infrared dark cloud labelled ``Dark Filament $\Jsto$ north'' in Figure\,\ref{fig:infrared} is the only CS(1-0) detection in the line of sight. The gas is foreground to much of the line-of-sight infrared emission. CO(1-0) emission at the same velocity is more spatially extended than CS(1-0) and $^{13}$CO(1-0), but also peaks in the northern region of $\Jsto$ corresponding to peaks of the tracers of the denser gas.


This campaign did not detect SNR-shock triggered thermal SiO(1-0) emission above an antenna sensitivity of 1$\sigma\sim$0.8\,K, and no thermal X-ray emission has been detected towards $\Jsto$. SiO(1-0) is a shock-tracer and the detection of this would have been `smoking gun' for a shock interaction with molecular gas. 
With the confirmed $\sim$3.2\,kpc kinematic distance for $\Jsto$, calculations employing this kinematic distance to yield a SNR Sedov age of 2-6$\times$10$^3$\,yr \citep{Acero:2015,Cui:2016} are increasingly likely, making $\Jsto$ potentially one of the older objects in the category of TeV shell morphology SNRs (RX\,J1713.7$-$3946, RX\,J0852$-$4622, RCW\,86 and SN\,1006 are 1.6, 2-4, 1.8 and 1\,kyr, respectively). 

On the `red-side' of the CO-traced Scutum-Crux arm (i.e. $\vlsr\sim -$10 to $-$5$\kms$), a void in molecular gas may exist towards $\Jsto$ (see bottom-left image of Figure\,\ref{fig:mom0s_2}), while a dense gas clump to the SNR's north-west is traced by $^{13}$CO and CS emission. We considered the possibility that this is the wind-blown bubble of the $\Jsto$ progenitor into which the supernova remnant expanded, but it is difficult to resolve this scenario with our other lines of evidence favouring the $\sim$3.2\,kpc kinematic distance solution for $\Jsto$ (i.e. X-ray absorption analysis), because this void is probably on the near-side edge of the Scutum-Crux arm. A proportion of the Scutum-Crux arm gas is required to be the X-ray photon-absorbing column density, so local intra-arm gas motions of 5-10\,$\kms$ would need to be invoked to reconcile this molecular void scenario with the other lines of evidence for the $\sim$3.2\,kpc kinematic distance. Alternatively, a systematic shift in the calculated column density, as discussed in Appendix\,\ref{Sec:app_distance} might allow the CO void scenario to be resolved with X-ray absorption measurements.

\subsubsection{CH$_3$OH emission}\label{sec:Methanol}
Locations of CH$_3$OH(7-6) emission \citep[][ and this paper]{Maxted:2015} are indicated in Figure\,\ref{fig:mom0s_2} and all correspond to dense gas traced by CO/$^{13}$CO/CS(1-0) emission. So-called `class I' CH$_3$OH masers are commonly associated with the molecular shocks of HII regions \citep[e.g.][]{Haschick:1990}, but in some cases have been suggested to be triggered by molecular shocks associated with old ($\sim$10$^4$\,yr) supernova remnants \citep[e.g. W28,][]{Nicholas:2012}. The CH$_3$OH detections in this survey are unlikely to be triggered by $\Jsto$ due to their spatial separation from the $\Jsto$ shock, although a connection with the C-shock is unclear for W28, where indirect ionisation by CRs may play a role \citep[e.g.][]{Maxted:2016b}.
The CH$_3$OH masers towards $\Jsto$ do illustrate the star-formation activity within the Scutum-Crux arm molecular gas that is traced by CO/$^{13}$CO/CS(1-0) in this study. 
CH$_3$OH lines are parametrised in Appendix\,\ref{sec:app_lines}, alongside other detected emission lines.

\subsubsection{The molecular mass towards $\Jsto$}\label{sec:MassScut}
The calculated molecular hydrogen column density peaks at $\sim$8$\times$10$^{23}$\,$\cmsqr$ (as derived from CS) within the Scutum-Crux arm at the approximate location ([$l,b$]$\sim$[353.41,$-$0.36]) of the HII region G353.43$-$0.37 (see Figure\,\ref{fig:ColDens}). CS(1-0) and $^{13}$CO(1-0) optical depth reach $\sim$5 and $\sim$0.6, respectively, towards this clump.

We estimate that the Scutum-Crux arm gas featured in Figure\,\ref{fig:mom0s_2} surveyed in this study has a total molecular mass of 9.3-16$\times 10^4$\,M$_{\odot}$, as traced by CO and $^{13}$CO. 
Approximately 4.3-5.1$\times 10^4$\,M$_{\odot}$ of Scutum-Crux arm molecular gas is traced towards $\Jsto$. 

An implication of the gas study by \citet{Fukuda:2014} was the attribution of most of the gamma-ray emission to hadronic processes in a region containing a mass of $\sim$6.4$\times$10$^4$\,$\Msun$ at a kinematic distance of 5.2\,kpc. A simple distance-squared-scaling of this analysis to the distance of the Scutum-Crux Arm gas (3.2\,kpc) would imply that only a few$\times$10$^4$\,M$_{\odot}$ would be required for an analogous hadronic scenario in the Scutum-Crux Arm. Generally though, a significant leptonic component for $\Jsto$ is expected \citep[see e.g.][]{Acero:2015,Cui:2016}, even in hadron-dominated models \citep[$\sim$20\%][]{Fukuda:2014}. 

If the $\sim$3.2\,kpc kinematic distance solution (discussed in previous sections) for $\Jsto$ holds, the SNR would be at a similar kinematic distance to the HII region G353.43$-$0.37. CS(1-0) emission reveals a dense cloud component associated with this object, implying an available mass of target material of order $\sim$10$^{5}$\,$M_{\odot}$. The new maps of diffuse and dense molecular gas will help refine future CR diffusion models in the context of a run-away CR origin for $\Jstn$ in slow and fast diffusion scenarios \citep{Cui:2016}. If the G353.43$-$0.37 cloud is penetrated by `runaway' CRs accelerated by $\Jsto$, it would be illuminated by gamma-ray emission, as is suggested by new preliminary HESS gamma-ray analyses \citep{Capasso:2016}. We note that dense gas can harbour strong frozen-in magnetic fields \citep[e.g.][]{Crutcher:2010} that can inhibit CR diffusion \citep[see e.g.][]{Gabici:2009}. As a result, dense clumps embedded within a SNR shell can anti-correlate with X-ray emission as high energy electrons emit non-thermally at clump boundaries \citep[e.g.][]{Sano:2010,Inoue:2012}. Further detailed investigation of such X-ray and ISM correlation with our Mopra data is left for future work.

\subsection{Gas in the 3\,kpc-Expanding Arm}\label{sec:3kpcArm}
Gas of the 3\,kpc-Expanding Arm ($\sim$5.2\,kpc) has been argued as candidate for association with $\Jsto$ (see Sections \ref{sec:intro} and \ref{sec:Distance}, and \citealt{Fukuda:2014}), however our Mopra CO(1-0) analyses (see Section\,\ref{sec:Xray}) appear to favour the foreground Scutum-Crux arm ($\sim$3.2\,kpc). 

If $\Jsto$ were in the 3kpc-Expanding arm, Mopra CO(1-0) measurements indicate that the molecular mass associated with this source would be $\sim$5.5$\times$10$^4$\,$\Msun$. 
A lack of CS(1-0) detection and a $^{13}$CO-derived mass of $\sim$20\% the CO mass suggests that the molecular gas is relatively diffuse, but the gas does extend south along the entire eastern edge of the SNR, and across the SNR's northern rim. This is consistent with the structure seen by the Nanten2 telescope \citep{Fukuda:2014}.

\subsection{Gas in the Norma-Cygnus Arm}\label{sec:NormaArm}
Approximately $\sim$8-11$\times$10$^4$\,$\Msun$ of Norma-Cygnus arm molecular gas was traced in this survey. 
No component is seen to overlap $\Jsto$. CO, $^{13}$CO and CS(1-0) emission is coincident with the HII region GAL\,353.56$-$00.01 \citep{Lockman:1989} at a consistent velocity ($\sim$57$\kms$). On close inspection, the CS(1-0) spectrum towards this region has a dual peak structure with peak velocities at, $\vlsr=-53.9\pm0.2$\,$\kms$ and $\vlsr=-58.4\pm0.1$\,$\kms$.

CO/$^{13}$CO/CS(1-0) emission also traces a clump at [$l,b$]$=$[353.40,$-$0.08] that has a coincident HII region GAL\,353.38$-$00.11 \citep[$\vlsr \sim 58$\,$\kms$, ][]{Lockman:1989}. Furthermore, the Extended Green Object, EGO\,G353.40$-$0.07 \citep{Chen:2013}, may indicate the presence of an outflow. The aforementioned infrared sources and the existence of several coincident young stellar objects \citep{Robitaille:2008} all point towards the dense CS(1-0)-traced gas of the Norma-Cygnus arm being an active region of star formation.

\section{Conclusions}
We surveyed the gamma-ray supernova remnant $\Jsto$ and nearby unidentified gamma-ray source $\Jstn$ with the Mopra radio telescope, targeting CO(1-0), $^{13}$CO(1-0) and CS(1-0) emission. We aimed to identify components of gas associated with $\Jsto$, and investigate the nature of $\Jstn$. We found that:
   \begin{enumerate}
      \item when adopting a CO X-factor towards the mid-range of published values, X-ray absorption column densities derived from the $\Jsto$ X-ray emission are consistent with column densities foreground to the Scutum-Crux arm at a line of sight velocity of $\sim -$15\,$\kms$, suggesting a kinematic distance of $\sim$3.2\,kpc for $\Jsto$.
      \item Components of dense molecular gas at $\sim$3.2\,kpc are coincident with the north of $\Jsto$, $\Jstn$ and a cloud associated with the HII region G353.43$-$0.37, as evidenced by CS(1-0) emission and infrared-dark features. The detection of dense gas towards gamma-ray emission to the north of $\Jsto$ is suggestive of a runaway CR scenario and flags a new component of target material mass to be included in future particle propagation models.
   \end{enumerate}

\section*{Acknowledgements}
The Mopra Telescope is part of the Australia Telescope and is funded by the Commonwealth of Australia for operation as a National Facility managed by the CSIRO. The University of New South Wales Mopra Spectrometer Digital Filter Bank used for these Mopra observations was provided with support from the Australian Research Council, together with the University of New South Wales, University of Sydney, Monash University and the CSIRO. We thank the Australian Research Council for helping to fund this work through a Linkage Infrastructure, Equipment and Facilities (LIEF) grant (LE16010094).




\bibliographystyle{mnras}
\bibliography{ReferencesHESSJ1731}





\appendix

\section{Systematic Uncertainty}\label{Sec:app_distance}


A systematic calibration error in either X-ray absorption column density, $N_{p}^{X}$, or spectral column density, $N_{p}^{CO+H}$, would shift the points of intercepts in Figures\,\ref{fig:XraySpecs1} and \ref{fig:XraySpecs2}, which would result in a shifted line-of-sight velocity solution for $\Jsto$ in our analysis.

\begin{figure}
\includegraphics[width=0.49\textwidth, angle=0]{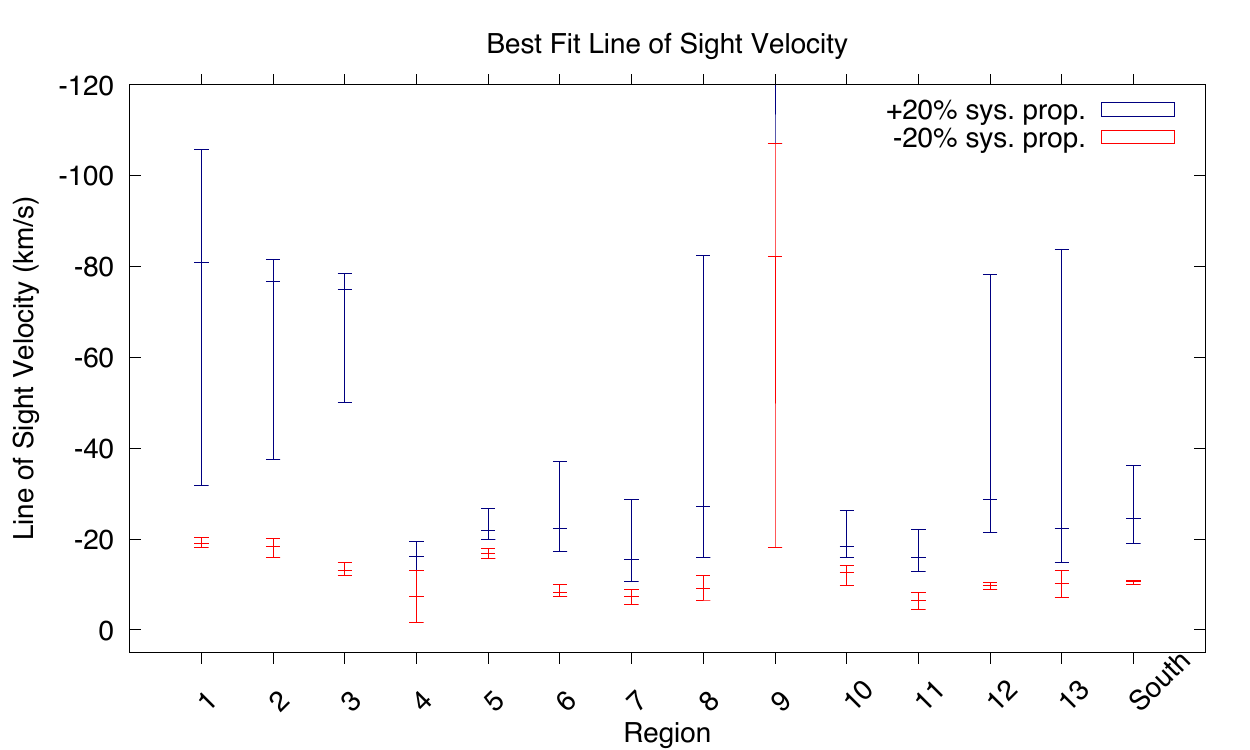}\\
\caption{The velocity of the intercepts of X-ray absorption column densities and corresponding spectrally-derived column densities for 14 regions towards the $\Jsto$ X-ray emission in the case of a 20\% systematic error. Thin navy and red error bars indicate 90\% confidence velocities that would be derived if X-ray absorption measurements or spectral column densities were affected by an overlooked systematic error of $\pm$20\%. The inverse variance-weighted mean of the two cases were $\vlsr = -$11.8$\pm$0.2\,$\kms$ and $\vlsr = -$20$\pm$1\,$\kms$.
\label{fig:BoxAndWhiskers_sys}}
\end{figure}

\citet{Bamba:2012} discuss the potential effect of a $\sim$20\% mis-estimation of spectral parameters in low surface-brightness regions, and conclude that the error introduced is smaller than statistical uncertainties, which itself is $\sim$few-20\% in most cases. We conduct an analysis of the potential effect of a systematic error in $N_{p}^{X}$ by repeating our analysis on a data-set with an artificially-injected offset of $\pm$20\%. We find that an over/underestimation of $N_{p}^{X}$ by 20\% would have little effect on the derived Scutum-Crux arm-association (preferred velocities of $-$11.8$\pm$0.2 and $-$20$\pm$1\,$\kms$ for 20\% offsets, see\,Figure\,\ref{fig:BoxAndWhiskers_sys}).

A similar argument can be made for the spectral column density, $N_{p}^{CO+H}$. The existence of an overestimation of $N_{p}^{CO+H}$ by up to $\sim$20\% would not change the derived Scutum-Crux arm-association, while a larger than 20\% underestimation can be tolerated to still yield the same results. 

Studies of HI disfavour a significant overestimation of atomic column density resulting from the assumed X-factor, but point towards the plausibility of underestimation. For example, cold HI cores contain a large component of mass not accounted for in an analysis of HI-emission alone due to self absorption \citep[e.g.][]{Krco:2010,Fukui:2012,Dickey:2013b}. Indeed, HI absorption features are apparent in the Scutum-Crux arm in several regions in Figures \ref{fig:XraySpecs1} and \ref{fig:XraySpecs2} (e.g. see Region\,1 at $-$17\,$\kms$ and Region\,2 at 0\,$\kms$ for clear cases), so an underestimation of atomic column density is possible in this analysis.

\citet{Strong:2004} used gamma-ray emission from hadronic interactions to estimate the CO X-factor applied in this analysis, and this CO X-factor value falls roughly in the mid-range of published values. A review by \citet{Bolatto:2013} suggests that the average CO X-factor is $2.0 \times 10^{20}\mathrm{cm}^{-2}.\mathrm{(K.km.s}^{-1})^{-1}$ with an uncertainty of 30\%, which suggests that an overestimation of molecular column density in this study is unlikely. We note that recent measurements by \citet{Okamoto:2017} suggest that the CO X-factor is $\sim$1$\times 10^{20}\mathrm{cm}^{-2}.\mathrm{(K.km.s}^{-1})^{-1}$ in the Perseus cloud. If such a value was valid for the region foreground to $\Jsto$, our X-ray analysis would be unable to discern between competing distance solutions.

Either an upwards shift in spectral column density, $N_{p}^{CO+H}$, or a downwards shift X-ray absorption column density, $N_{p}^{X}$, would result in the preferred line of sight velocity shifting towards the near-side of the Scutum-Crux arm. A very large shift ($\sim$60\%) in these directions would be required to move the intercepts in Figures\,\ref{fig:XraySpecs1} and \ref{fig:XraySpecs2} towards velocities corresponding to the Sagittarius arm ($\sim$0$\kms$, $\sim$1\,kpc).

\begin{figure}
\includegraphics[width=0.49\textwidth]{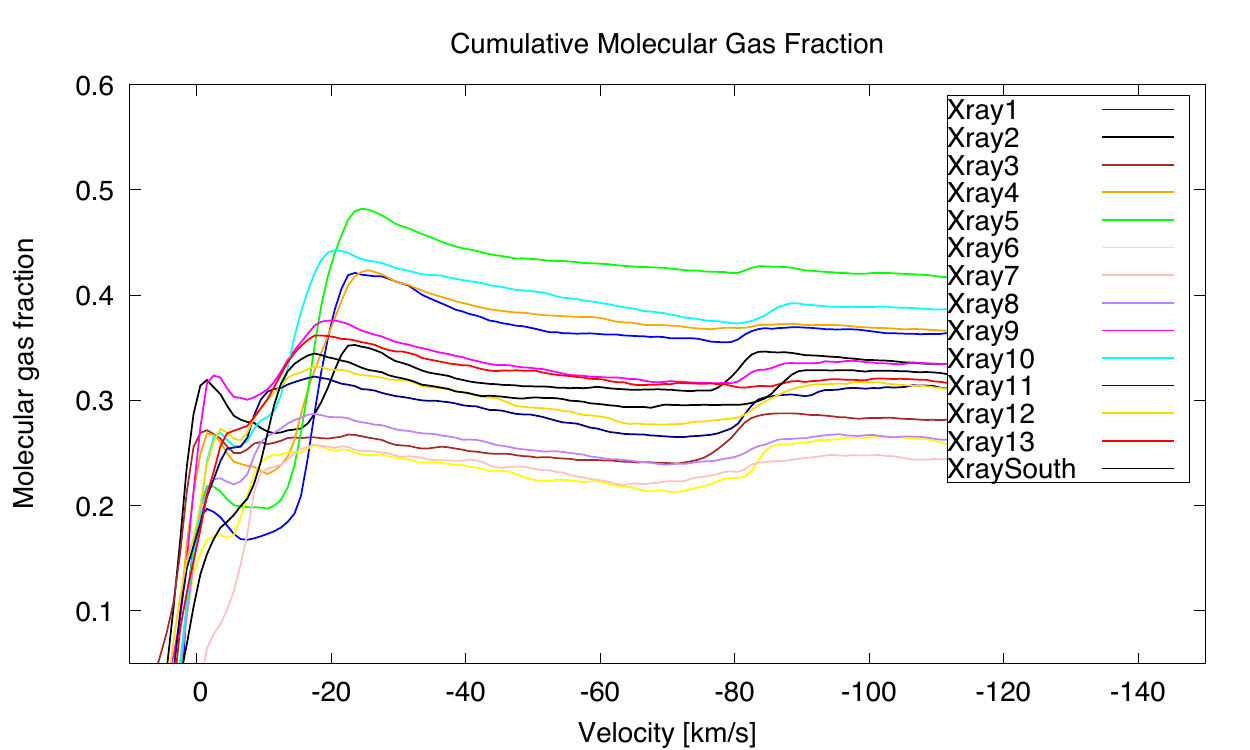}
\caption{Molecular gas fractions ([H$_2$]/[H+H$_2$]) towards $\Jsto$.}\label{fig:MolGasFrac}
\end{figure}

We note that the results of an investigation of the molecular gas fraction towards $\Jsto$ X-ray regions are presented in Figure\,\ref{fig:MolGasFrac}. The ratio of molecular to atomic gas is observed to be within normal ranges (([H$_2$]/[H+H$_2$])$\sim$0.2-0.4), illustrating that it is important to include both atomic and molecular column density contributions.
Figure\,\ref{fig:MolGasFrac} illustrates the molecular gas fraction towards 14 regions of $\Jsto$. Values are consistent with other studies \citep[$>$0.25,0.4-0.45,$\sim$0.35,0.38-0.45][]{Bohlin:1978,Liszt:2002,Liszt:2010,Maxted:2013ctb}.

\section{Infrared-dark clouds towards $\Jsto$ and $\Jstn$}\label{sec:app_lines}
\begin{figure}
\includegraphics[width=0.48\textwidth, angle=0]{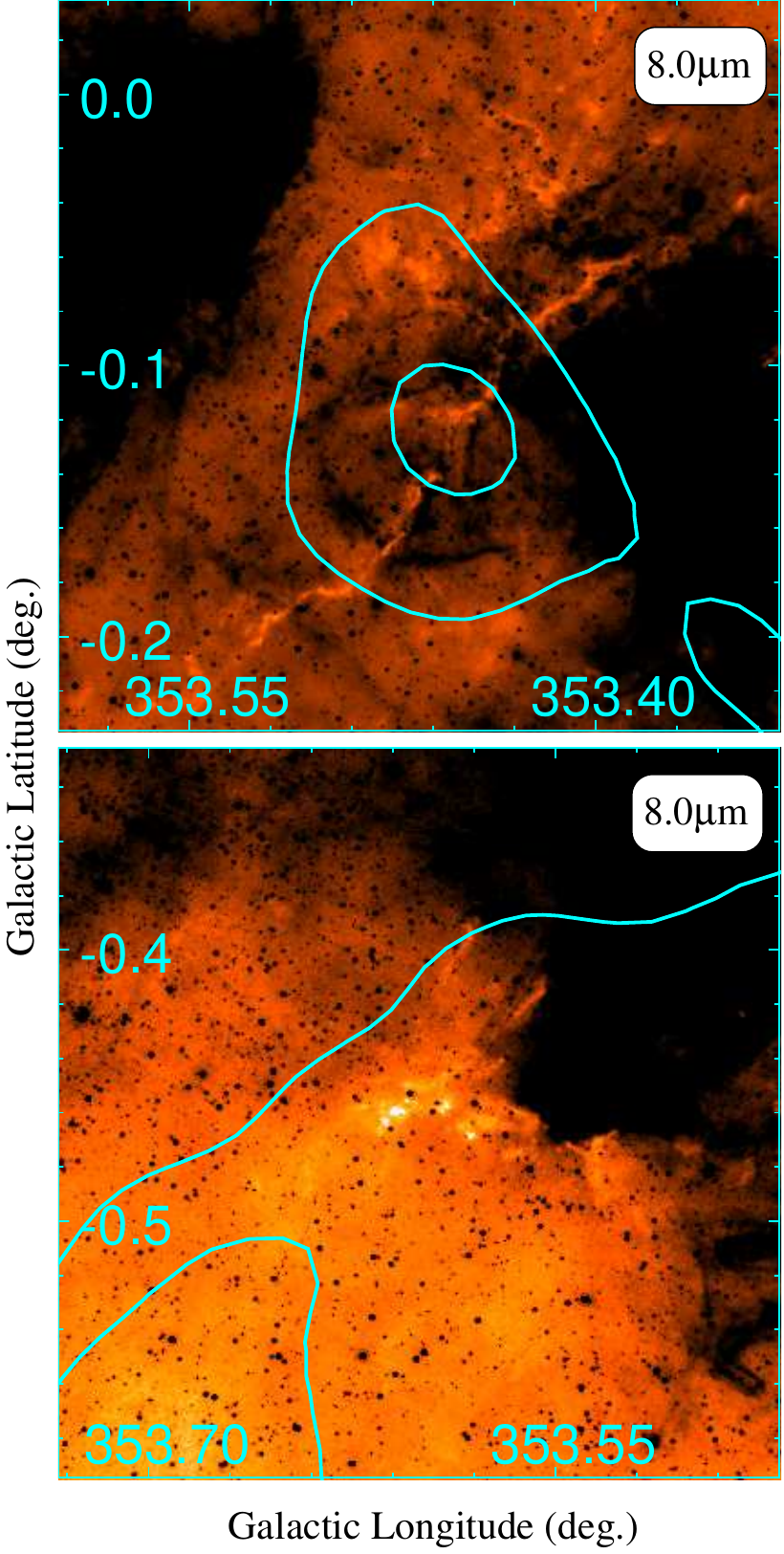}
\caption{8\,$\mu$m images of the dark filiment towards $\Jstn$  (top) and the ``Dark Filament $\Jsto$ north'' (bottom) \citep[GLIMPSE,][]{Carey:2009}. Dark and light regions correspond to high and low-intensity regions, respectively. The colour scales have been adjusted to emphasize infrared-dark gas. HESS 4, 6 and 8$\sigma$ $>$1\,TeV gamma-ray significance contours (thick cyan) are overlaid \citep{Abramowski:2011}.}\label{fig:infraredZoom}
\end{figure}

In Section\,\ref{sec:ScutArm}, infrared-dark features were shown to be coincident with dense gas in the Scutum-Crux arm, as traced by CS(1-0) and $^{13}$CO emission. In addition to Figure\,\ref{fig:infrared}, we provide zoomed-in 8\,$\mu$m images towards $\Jsto$ and $\Jstn$ in Figure\,\ref{fig:infraredZoom}. No additional molecular tracers were seen towards these 2 infrared dark clumps. As described in the main body of this paper, detections of other molecular gas tracers SiO, CH$_3$OH and HC$_3$N are summarised in Table\,\ref{tab:LineParams}. This data will be made publicly-available alongside CS(1-0) data (see Section\,\ref{sec:CS}).

\begin{table}
\centering
\caption{Gaussian function fit parameters for CH$_3$OH(7$_0$-6$_1$\,A$++$) emission at 44.06947\,GHz and SiO(1-0,v=2) emission at 42.820582\,GHz, HC$_3$N(5-4) emission at 45.488839\,GHz, $^{13}$CS(1-0) emission at 46.247580\,GHz and C$^{34}$S(1-0) emission at 48.206946\,GHz. The fitted functions were of the form $\Tpeak.\exp(-(v-\vlsr)^2/(2.\textrm{v}_{\sigma}^2))$. 
\label{tab:LineParams}} 
\small
\begin{tabular}{|l|r|r|r|}
\hline
Source 					&	$\vlsr$		&	$\Tpeak$		&	v$_{\sigma}$ \\
$[l,b]$					&	[$\kms$]		&	[K]			&	[$\kms$]\\
\hline
\textbf{CH$_3$OH(7$_0$-6$_1$\,\tiny{A++})} & & &\\
A [353.21,$-$0.24]	&	$-$15.40$\pm$0.03&	0.43	$\pm$0.06	&	0.17$\pm$0.03\\
A 					&	5.92$\pm$0.05	&	0.22$\pm$0.05	&	0.13	$\pm$0.04\\
B [353.25,$-$0.50]	&	$-$8.91$\pm$	0.02	&	0.54$\pm$0.05	&	0.17$\pm$0.02\\
C [353.34,$-$0.32]	&	$-$17.97$\pm$0.02&	0.37$\pm$0.03	&	0.22$\pm$0.02\\
D [353.41,$-$0.37]	&$-$16.22$\pm$0.02&	0.66$\pm$0.03	&	0.27$\pm$0.02\\
D 					&$-$17.73$\pm$0.10&	0.33$\pm$0.01	&	1.99$\pm$0.08	\\
\hline
\textbf{SiO(1-0,v=0)} & & & \\
D & $-$17.0$\pm$0.1	&	0.243$\pm$0.007 &	3.3$\pm$0.1 \\
\hline
\textbf{HC$_3$N(5-4)} 	& & & \\
D & $-$16.9$\pm$0.1		& 0.50$\pm$0.02 	&	1.72$\pm$0.09 \\
D & $-$18.74$\pm$0.06	& 0.28$\pm$0.04	& 	0.45$\pm$0.08 \\
\hline
\textbf{$^{13}$CS(1-0)} & & & \\
D & $-$16.7$\pm$0.2 &0.099$\pm$0.008 &	1.9$\pm$0.2 \\
\hline
\textbf{C$^{34}$S(1-0)} & & & \\
D &	-16.91$\pm$0.09 &	0.27$\pm$0.01 &	1.94$\pm$0.09 \\
\hline
\textbf{CS(1-0)} & & & \\
D &	-17.05$\pm$0.04 &	1.67$\pm$0.02 &	2.48$\pm$0.05 \\
\hline
\textbf{SiO(1-0,v=2)} & & & \\
E [353.61,$-$0.24]		&$-$85.41$\pm$0.07&	0.30$\pm$0.02	&	1.13$\pm$0.07	\\
\hline
\end{tabular}
\end{table}

\end{document}